\documentclass[a4paper,11pt]{article}
\usepackage{jinstpub} 
\usepackage[mathlines]{lineno}

\usepackage{graphicx}
\usepackage[utf8]{inputenc}
\usepackage{xcolor}
\usepackage{xspace}
\usepackage{ulem} 
\usepackage{hyperref}
\usepackage{amsmath}
\usepackage{accents}
\usepackage{placeins}

\graphicspath{{figures/}}

\newcommand{\numu}{\ensuremath{\nu_{\mu}}\xspace}

\newcommand{\sinsq}[1]{\ensuremath{\sin^2 \theta_{#1}}\xspace}
\newcommand{\dmsq}[1]{\ensuremath{\Delta m^2_{#1}}\xspace}
\newcommand{\logl}{\ensuremath{-2 \ln\! \mathcal{L}}\xspace}
\newcommand{\logll}{\ensuremath{-2 \ln\! \mathcal{L}/\mathcal{L}_{0}}\xspace}

\newcommand{\LR}{\ensuremath{\lambda}\xspace}
\newcommand{\LL}{\ensuremath{\ell}\xspace}

\newcommand{\crit}{\ensuremath{c_\alpha}\xspace}
\newcommand{\criti}{\ensuremath{c_{\alpha,i}}\xspace}
\newcommand{\critw}{\ensuremath{c_{\alpha,\textrm{Wilks}}}\xspace}
\newcommand{\chisq}{\ensuremath{\chi^{2}}\xspace}

\newcommand{\dcp}{\ensuremath{\delta_{\textrm{CP}}}\xspace}

\newcommand{\hatdelta}{\ensuremath{\hat\delta}\xspace}
\newcommand{\thvec}{\ensuremath{\boldsymbol{\theta}}\xspace}
\newcommand{\xvec}{\ensuremath{\boldsymbol{x}}\xspace}
\newcommand{\nuisvec}{\ensuremath{\boldsymbol{\phi}}\xspace}

\newcommand{\nexp}{\ensuremath{N_{\rm exp}}\xspace}
\newcommand{\nobs}{\ensuremath{N_{\rm obs}}\xspace}
\newcommand{\thetamix}{\ensuremath{\theta_{23}}\xspace}
\newcommand{\CLs}{\texorpdfstring{CL$_\text{S}$}{CLs}\xspace}
\newcommand{\term}[1]{`#1'}
\newcommand{\Ndata}{\ensuremath{N_{\textrm{data}}}}
\newcommand{\sigst}{\ensuremath{\sigma_{\textrm{stat}}}}
\newcommand{\sigsy}{\ensuremath{\sigma_{\textrm{syst}}}}

\DeclareMathOperator*{\argmin}{argmin}
\DeclareMathOperator*{\argmax}{argmax}
\DeclareMathOperator*{\erfc}{erfc}

\newlength{\dhatheight}
\newcommand{\doublehat}[1]{%
    \settoheight{\dhatheight}{\ensuremath{\hat{#1}}}%
    \addtolength{\dhatheight}{-0.2ex}%
    \hat{\vphantom{\rule{1pt}{\dhatheight}}%
    \smash{\hat{#1}}}}

\newcommand{\fig}[1]{Figure~\ref{fig:#1}}

\newcommand{\eqn}[1]{Equation~\ref{eqn:#1}}
\newcommand{\sect}[1]{Section~\ref{sec:#1}}

\definecolor{darkgreen}{rgb}{0, .7, 0}

\title{Monte Carlo method for constructing confidence intervals with unconstrained and constrained nuisance parameters in the NOvA experiment}

\newcommand{\ANL}{Argonne National Laboratory, Argonne, Illinois 60439, 
USA}
\newcommand{\ICS}{Institute of Computer Science, The Czech 
Academy of Sciences, 
182 07 Prague, Czech Republic}
\newcommand{\IOP}{Institute of Physics, The Czech 
Academy of Sciences, 
182 21 Prague, Czech Republic}
\newcommand{\Atlantico}{Universidad del Atlantico,
Carrera 30 No. 8-49, Puerto Colombia, Atlantico, Colombia}
\newcommand{\BHU}{Department of Physics, Institute of Science, Banaras 
Hindu University, Varanasi, 221 005, India}

\newcommand{\Caltech}{California Institute of 
Technology, Pasadena, California 91125, USA}
\newcommand{\Cochin}{Department of Physics, Cochin University
of Science and Technology, Kochi 682 022, India}
\newcommand{\Charles}
{Charles University, Faculty of Mathematics and Physics,
 Institute of Particle and Nuclear Physics, Prague, Czech Republic}
\newcommand{\Cincinnati}{Department of Physics, University of Cincinnati, 
Cincinnati, Ohio 45221, USA}
\newcommand{\CSU}{Department of Physics, Colorado 
State University, Fort Collins, CO 80523-1875, USA}
\newcommand{\CTU}{Czech Technical University in Prague,
Brehova 7, 115 19 Prague 1, Czech Republic}

\newcommand{\Delhi}{Department of Physics and Astrophysics, University of 
Delhi, Delhi 110007, India}
\newcommand{\JINR}{Joint Institute for Nuclear Research,  
Dubna, Moscow region 141980, Russia}
\newcommand{\Erciyes}{
Department of Physics, Erciyes University, Kayseri 38030, Turkey}
\newcommand{\FNAL}{Fermi National Accelerator Laboratory, Batavia, 
Illinois 60510, USA}
\newcommand{\UFG}{Instituto de F\'{i}sica, Universidade Federal de 
Goi\'{a}s, Goi\^{a}nia, Goi\'{a}s, 74690-900, Brazil}
\newcommand{\Guwahati}{Department of Physics, IIT Guwahati, Guwahati, 781 
039, India}
\newcommand{\Harvard}{Department of Physics, Harvard University, 
Cambridge, Massachusetts 02138, USA}
\newcommand{\Houston}{Department of Physics, 
University of Houston, Houston, Texas 77204, USA}
\newcommand{\IHyderabad}{Department of Physics, IIT Hyderabad, Hyderabad, 
502 205, India}
\newcommand{\Hyderabad}{School of Physics, University of Hyderabad, 
Hyderabad, 500 046, India}
\newcommand{\IIT}{Illinois Institute of Technology,
Chicago IL 60616, USA}
\newcommand{\Indiana}{Indiana University, Bloomington, Indiana 47405, 
USA}
\newcommand{\INR}{Institute for Nuclear Research of Russia, Academy of 
Sciences 7a, 60th October Anniversary prospect, Moscow 117312, Russia}
\newcommand{\Iowa}{Department of Physics and Astronomy, Iowa State 
University, Ames, Iowa 50011, USA}
\newcommand{\Irvine}{Department of Physics and Astronomy, 
University of California at Irvine, Irvine, California 92697, USA}

\newcommand{\Lebedev}{Nuclear Physics and Astrophysics Division, Lebedev 
Physical 
Institute, Leninsky Prospect 53, 119991 Moscow, Russia}
\newcommand{\Magdalena}{Universidad del Magdalena, Carrera 32 No 22-08 Santa Marta, Colombia}
\newcommand{\MSU}{Department of Physics and Astronomy, Michigan State 
University, East Lansing, Michigan 48824, USA}

\newcommand{\Duluth}{Department of Physics and Astronomy, 
University of Minnesota Duluth, Duluth, Minnesota 55812, USA}
\newcommand{\Minnesota}{School of Physics and Astronomy, University of 
Minnesota Twin Cities, Minneapolis, Minnesota 55455, USA}
\newcommand{\Mississippi}{University of Mississippi, University, Mississippi 38677, USA}
\newcommand{\NISER}{National Institute of Science Education and Research,
Khurda, 752050, Odisha, India}

\newcommand{\Panjab}{Department of Physics, Panjab University, 
Chandigarh, 160 014, India}
\newcommand{\Pitt}{Department of Physics, 
University of Pittsburgh, Pittsburgh, Pennsylvania 15260, USA}
\newcommand{\QMU}{Particle Physics Research Centre, 
Department of Physics and Astronomy,
Queen Mary University of London,
London E1 4NS, United Kingdom}

\newcommand{\SAlabama}{Department of Physics, University of 
South Alabama, Mobile, Alabama 36688, USA} 
\newcommand{\Carolina}{Department of Physics and Astronomy, University of 
South Carolina, Columbia, South Carolina 29208, USA}

\newcommand{\SMU}{Department of Physics, Southern Methodist University, 
Dallas, Texas 75275, USA}

\newcommand{\Sussex}{Department of Physics and Astronomy, University of 
Sussex, Falmer, Brighton BN1 9QH, United Kingdom}
\newcommand{\Syracuse}{Department of Physics, Syracuse University,
Syracuse NY 13210, USA}

\newcommand{\Texas}{Department of Physics, University of Texas at Austin, 
Austin, Texas 78712, USA}
\newcommand{\Tufts}{Department of Physics and Astronomy, Tufts University, Medford, 
Massachusetts 02155, USA}
\newcommand{\UCL}{Physics and Astronomy Department, University College 
London, 
Gower Street, London WC1E 6BT, United Kingdom}
\newcommand{\Virginia}{Department of Physics, University of Virginia, 
Charlottesville, Virginia 22904, USA}
\newcommand{\WSU}{Department of Mathematics, Statistics, and Physics,
 Wichita State University, 
Wichita, Kansas 67206, USA}
\newcommand{\WandM}{Department of Physics, William \& Mary, 
Williamsburg, Virginia 23187, USA}
\newcommand{\Wisconsin}{Department of Physics, University of 
Wisconsin-Madison, Madison, Wisconsin 53706, USA}

\newcommand{\LAtlantico}{a}
\newcommand{\LMississippi}{b}
\newcommand{\LFNAL}{c}
\newcommand{\LJINR}{d}
\newcommand{\LMagdalena}{e}
\newcommand{\LSussex}{f}
\newcommand{\LCincinnati}{g}
\newcommand{\LIndiana}{h}
\newcommand{\LIowa}{i}
\newcommand{\LUCL}{j}
\newcommand{\LVirginia}{k}
\newcommand{\LIrvine}{l}
\newcommand{\LHyderabad}{m}
\newcommand{\LTufts}{n}
\newcommand{\LErciyes}{o}
\newcommand{\LCaltech}{p}
\newcommand{\LIIT}{q}
\newcommand{\LPanjab}{r}
\newcommand{\LGuwahati}{s}
\newcommand{\LMinnesota}{t}
\newcommand{\LQMU}{u}
\newcommand{\LIHyderabad}{v}
\newcommand{\LMSU}{w}
\newcommand{\LCSU}{x}
\newcommand{\LINR}{y}
\newcommand{\LTexas}{z}
\newcommand{\LWisconsin}{aa}
\newcommand{\LWandM}{ab}
\newcommand{\LDelhi}{ac}
\newcommand{\LSMU}{ad}
\newcommand{\LANL}{ae}
\newcommand{\LHarvard}{af}
\newcommand{\LIOP}{ag}
\newcommand{\LCTU}{ah}
\newcommand{\LSAlabama}{ai}
\newcommand{\LPitt}{aj}
\newcommand{\LUFG}{ak}
\newcommand{\LLebedev}{al}
\newcommand{\LCarolina}{am}
\newcommand{\LDuluth}{an}
\newcommand{\LICS}{ao}
\newcommand{\LHouston}{ap}
\newcommand{\LCochin}{aq}
\newcommand{\LWSU}{ar}
\newcommand{\LBHU}{as}
\newcommand{\LCharles}{at}
\newcommand{\LNISER}{au}
\newcommand{\LSyracuse}{av}

\author[\LAtlantico]{M.~A.~Acero}
\author[\LMississippi]{B.~Acharya}
\author[\LFNAL]{P.~Adamson}
\author[\LFNAL]{L.~Aliaga}
\author[\LJINR]{N.~Anfimov}
\author[\LJINR]{A.~Antoshkin}
\author[\LMagdalena]{E.~Arrieta-Diaz}
\author[\LSussex]{L.~Asquith}
\author[\LCincinnati]{A.~Aurisano}
\author[\LIndiana,\LIowa]{A.~Back}
% ...

\collaboration{The NOvA Collaboration}

\date{\today}

\emailAdd{ahimmel@fnal.gov}


\arxivnumber{2207.14353 } 

\abstract{Measuring observables to constrain models using maximum-likelihood estimation is fundamental to many physics experiments.
Wilks' theorem provides a simple way to construct confidence intervals on model parameters, but it only applies under certain conditions. These conditions, such as nested hypotheses and unbounded parameters, are often violated in neutrino oscillation measurements and other experimental scenarios. Monte Carlo methods can address these issues, albeit at increased computational cost. In the presence of nuisance parameters, however, the best way to implement a Monte Carlo method is ambiguous. 
This paper documents the method selected by the NOvA experiment, the profile construction. It presents the toy studies that informed the choice of method, details of its implementation, and tests performed to validate it. It also includes some practical considerations which may be of use to others choosing to use the profile construction. 
}

\begin{document}
\maketitle
\flushbottom

\section{Introduction}\label{sec:introduction}

The main goal of many physics experiments is to make measurements of the properties of Nature in the form of parameters of a model. Often, those parameters cannot be observed directly, and must instead be inferred from a likelihood function,  $\mathcal L(\xvec|\thvec)$, which describes the probability of the observed data, \xvec, for a given set of parameter values, $\thvec$. In frequentist analyses, the best estimate for the model parameters is determined using maximum likelihood estimation. Results are usually~\cite{ref:pdg} presented as one- or two-dimensional Neyman--constructed confidence intervals~\cite{ref:neymanconstruction}, and Wilks' theorem~\cite{ref:wilks} is used to determine the confidence level which corresponds to a given likelihood value. 
However, Wilks' theorem is only valid if certain conditions are met, so some experimental measurements that depend on Wilks' theorem may
fail to produce confidence intervals with reasonable frequentist \term{coverage,}  meaning that confidence intervals determined in the same way in many repeated experiments would not contain the true value with the reported frequency. In other words, the confidence intervals would have an actual significance different from what is reported. Monte Carlo methods with various implementations~\cite{ref:diggle,ref:kendall} have long been proposed as a solution to this issue. The Unified Approach~\cite{ref:fc} is a type of Monte Carlo method which defines a nonparametric ordering procedure for determining the critical values that define the extent of the confidence intervals. The method is commonly known in the high energy physics community as the \term{Feldman--Cousins} (FC) method, after the authors who popularized it in the field. 

However, the Feldman--Cousins paper does not give guidance on how to handle additional nuisance parameters beyond those being measured, making implementation ambiguous in experiments where nuisance parameters are present. Ensuring reasonable coverage in the presence of nuisance parameters is a challenge. No method can guarantee correct coverage for all possible values of the nuisance parameters, but various approaches can give more or less accurate coverage. 
This challenge is acute for long-baseline oscillation experiments like NOvA since the neutrino oscillation probabilities depend on three parameters values ($|\dmsq{32}|$, $\theta_{23}$, \dcp) and a binary choice on the sign of the \dmsq{32}, known as the neutrino mass ordering~\cite{ref:nova3f2017, ref:nova3f2018, ref:nova3f2019, ref:nova3f2021}. As a consequence, any confidence interval drawn in one or two dimensions will have both constrained nuisance parameters (systematic uncertainties) and unconstrained nuisance parameters (the other model parameters). 

This paper presents the technique used in the neutrino oscillation measurements made by the NOvA experiment~\cite{ref:nova3f2017, ref:nova3f2018, ref:nova3f2019, ref:nova3f2021}. \sect{methods} provides a brief pedagogical introduction to the Feldman-Cousins method and the challenge presented by nuisance parameters, as well as several possible approaches to the problem. \sect{toy} presents simplified toy models used to evaluate the different methods and select the \emph{profile construction}~\cite{ref:kendall,ref:phystat2003,ref:phystat2005,ref:phystat2007} for use in NOvA. \sect{nova} describes the implementation of this method in practice, including some methods used to validate its coverage, and important features of the confidence intervals it produces.


\section{Confidence Interval Construction with Nuisance Parameters}
\label{sec:methods}

\subsection{The Original Feldman--Cousins Method}\label{sec:orig_FC}

A common method for creating frequentist confidence intervals is the Neyman construction~\cite{ref:neymanconstruction}. Likelihood--ratio tests are performed between each point in parameter space and the best fit point, with test statistic $\LR$ defined as: 

\begin{linenomath*}
\begin{equation}\label{eqn:lrtest}
    \LR_i = -2 \ln\! \frac{\mathcal{L}(\xvec|\thvec_i)}{\mathcal{L}(\xvec|\hat\thvec)} = \LL(\xvec|\thvec_i)-\LL(\xvec|\hat\thvec),
\end{equation}
\end{linenomath*}
where $\mathcal L(\xvec|\thvec)$ is the likelihood function of data \xvec given parameter values \thvec, $\ell$ is $\logl$, $\thvec_{i}$ is the $i^\textrm{th}$ set of fixed values of the parameters being tested for potential inclusion in the confidence interval, and $\hat\thvec$ is the overall maximum likelihood estimate, hereinafter referred to as \term{best fit,} of all parameters to the data. Point $i$ is included in the $\alpha$-level confidence interval if the $p$-value from the likelihood ratio test is less than $1-\alpha$, or equivalently, if $\LR_i$ is less than a \term{critical value,} $c_\alpha$, given by:
\begin{linenomath*}
\begin{equation}\label{eqn:critval}
    \int_{0}^{c_\alpha} P(\LR_i) d\LR_i = \alpha,
\end{equation}
\end{linenomath*}
where $P$ is the expected distribution of the $\lambda_i$ statistic assuming the true $\thvec=\thvec_i$.
As can be seen from \eqn{critval}, calculating the critical value requires knowledge of the distribution of the likelihood--ratio test statistic.

If the conditions of Wilks' theorem \cite{ref:wilks} are met, then the distribution $P(\LR)$ asymptotically approaches a \chisq distribution with a number of degrees of freedom equal to the number of parameters of interest\footnote{The number of parameters of interest is equivalent to the difference in number of degrees-of-freedom between the two likelihoods in the likelihood ratio.} with deviations\footnote{In practice, these deviations are quite small even for small $N$ when the data is Poisson-distributed~\cite{ref:cordeiro}.} expected at the $\mathcal{O}(1/\sqrt{N})$ level, where $N$ refers to the size of the data sample, \xvec. This asymptotic behavior means $P(\lambda)$ is the same for any point, $i$. Since the \chisq distributions are well known, fixed critical values for drawing confidence intervals at common significance levels are tabulated and readily available.

The conditions required for Wilks' theorem to apply are: (1) the maximum likelihood estimators of the parameters have ellipsoidal distributions, and (2) the null hypothesis is \term{nested} within the range of alternative hypotheses. A common way to violate assumption (1) is a physical boundary on the possible values of a parameter applied externally (e.g., probabilities must be between 0 and 1), but it can also be violated by an effective boundary introduced by a function with a limited range such as $\sin()$, or degeneracies that introduce other, potentially disjoint, regions of parameter values which are potentially consistent with the observed data.\footnote{Boundaries tend to reduce freedom to find optimal fits to the data and shrink confidence intervals, while degeneracies tend to add freedom and expand intervals, but in both cases the assumption of an ellipsoidal distribution is violated. In some circumstances, there are variations to Wilks’ theorem that can still give asymptotic distributions in the presence of boundaries \cite{Cowan:2010js}. }.  
The specific ways the NOvA oscillation measurement violates these assumptions is explained in more detail in Section~\ref{sec:nova_wilks}.
When the assumptions of Wilks' theorem are not satisfied, the significance of the hypothesis tests cannot be reliably determined using the \chisq distribution, meaning this method will not produce correct coverage for confidence intervals at
their reported significance -- another method must be used to determine suitable critical values. 

The Feldman--Cousins (FC) method~\cite{ref:fc} provides a nonparametric approach to defining confidence intervals with correct coverage and is commonly used in particle physics. A large number, $N$, of FC pseudoexperiments are simulated at points sampling the range of parameter values where confidence intervals will be reported. A \term{Feldman--Cousins pseudoexperiment} represents a possible experimental observation at a given set of parameters, \thvec. Each pseudoexperiment is constructed by drawing a Poisson-distributed random number for each bin of our analysis samples, with the mean of those Poisson distributions being the predicted number of events in that bin given \thvec. For each FC pseudoexperiment, $\xvec_j$, the best fit of the parameter(s), $\hat\thvec_{j}$, is also found through Maximum Likelihood Estimation. The FC pseudoexperiments are then ordered by the difference in \LL between the \term{true} value used to generate the FC pseudoexperiments and the best fit, 
\begin{linenomath*}
\begin{equation}
    \LR_{ij} = \LL(\xvec_j|\thvec_{i}) - \LL(\xvec_j|\hat\thvec_{j}),
\end{equation}
\end{linenomath*}
to form a distribution $P(\LR_i)$ that differs for every $\thvec_i$. This procedure is called \term{nonparametric} since the ordering of the pseudoexperiments creates a distribution for the test statistic, $\LR_i$, without knowing in advance how it should be distributed.
Then, the $\alpha$-significance-level critical value for this set of true parameters, $\crit(\thvec_i)$ as defined in \eqn{critval}, is the value which is larger than the first $\alpha N$ of the $\LR_{ij}$ values. This procedure is then repeated for each point being tested, and the confidence interval at level $\alpha$ is made up of the points where $\LR_i < \crit(\thvec_i)$.
If the FC pseudoexperiments are a fair representation of the data, it is straightforward to see that this procedure will give correct coverage, $\alpha$, since we have empirically determined for each point in parameter space the critical value $\crit(\thvec_i)$ which will cover $\alpha$ fraction of the pseudoexperiments generated with values $\thvec_i$.

\subsection{The Challenge of Nuisance Parameters}

While the above procedure from~\cite{ref:fc} is straightforward, it does not provide guidance on a key question when applying it in practice: how to handle nuisance parameters. We use the term \term{nuisance parameters} (hereinafter referred to by \nuisvec to distinguish them from the parameters of interest, \thvec) to refer to any model parameter that we do not wish to include in the specification of our final confidence intervals. These can be `physics' parameters the experiment is measuring, but whose constraints are not reported in a particular interval, other parameters of the model which are constrained by external experiments, or parameters representing systematic uncertainties, whose exact values are uninteresting. 

A common approach for handling nuisance parameters is to \term{profile} over them~\cite{ref:kendall}. That is, at each point in the parameter space, $\thvec_i$, at which the likelihood is to be evaluated, a search is performed over all values of the nuisance parameters, and the combination of nuisance parameters that yield the maximum likelihood (minimum $\ell$), 
\begin{linenomath*}
\begin{equation}\label{eqn:nuisance}
    \doublehat\nuisvec_i = \argmin_{\nuisvec}\ell(\thvec_i, \nuisvec),
\end{equation}
\end{linenomath*}
is adopted. $\doublehat\nuisvec_i$, which corresponds to point $\thvec_i$, is marked with two hats to distinguish it from the globally optimal nuisance parameters, $\hat\nuisvec$, which correspond to the best estimate of the parameters of interest, $\hat\thvec$. With these parameters defined, the likelihood ratio from \eqn{lrtest} becomes:
\begin{linenomath*}
\begin{equation}\label{eqn:lrtestprof}
    \LR_i = \LL(\xvec|\thvec_i, \doublehat{\nuisvec}_i)-\LL(\xvec|\hat\thvec, \hat\nuisvec).
\end{equation}
\end{linenomath*}
In the frequentist statistical philosophy each nuisance parameter possesses an (unknown) true value. The intuition is that, absent any further information, we adopt the nuisance parameter values most compatible with the data. This procedure contrasts with the Bayesian \term{marginalization} procedure, where the likelihood is taken to be the likelihood integrated over all values of the nuisance parameters, weighted by a prior probability distribution. 

The coverage guarantees of the Feldman--Cousins procedure rely on our access to a collection of FC pseudoexperiments to inspect, which have been generated at the precise points we wish to include/exclude at a certain significance. In the presence of nuisance parameters, however, we no longer have access to such an ensemble since the values of the nuisance parameters are not defined a priori by the point in parameter space being tested. Nevertheless, some values must be chosen in order to generate FC pseudoexperiments.
We could ensure correct coverage by defining our confidence intervals in a high-dimensional space containing all the nuisance parameters, but this is impractical, both computationally and because it cannot be easily visualized. When defining a lower-dimensional confidence interval, the values we choose for the nuisance parameters may differ from the true values, potentially yielding incorrect coverage. Note, the goal in choosing nuisance parameters for the pseudoexperiments is to ensure accurate coverage; it is not intended as a method for propagating systematic uncertainties to confidence intervals. That goal is accomplished by including them as nuisance parameters in the original likelihood.

\subsection{Approaches to Nuisance Parameters in Monte Carlo Methods}

Several plausible approaches exist for generating the FC pseudoexperiments for point $\thvec_i$ in the presence of nuisance parameters; the methods differ both in how practical they are to use and in the accuracy of the coverage they achieve. We discuss the methods below, and point out those which are impractical to apply to real-world problems. The coverage properties of the methods that are practical to implement will be explored in \sect{toy}.

\begin{description}
    \item[A priori estimate] Hold the nuisance parameters fixed at their a priori assumed values in the generation of all FC pseudoexperiments, $\nuisvec_i = \nuisvec_0$. While straightforward, in the plausible case that the true values of the nuisance parameters differ from their a priori values, the a priori estimate solution ignores the information available from the data about their values and thus can easily under- or over-cover. While not expected to perform well, this method is straightforward to implement so we will examine its coverage properties in \sect{toy}.

    \item[Conservative] At each point in the parameter space, $\thvec_i$, select the values of the nuisance parameters that yield the most conservative (largest) critical value based on FC pseudoexperiments, and thus the largest confidence interval, $\nuisvec_i = \argmax_{\nuisvec} \criti(\nuisvec)$. By taking the most conservative critical values, this method is guaranteed not to under-cover. However, because even nuisance parameters highly inconsistent with the data are considered, it is likely to substantially over-cover. Additionally, unless a closed-form estimate of the $\criti(\nuisvec)$ is available, this can be computationally infeasible for unbounded parameters or a large number of parameters.
    \item[Berger--Boos] This method is philosophically similar to the conservative method, but introduces a limiting principle for which values of nuisance parameter to consider. At each point in parameter space, $\thvec_i$, determine the range of nuisance parameters consistent with the data at significance level $\beta$, and then calculate $p$-values empirically (i.e. using pseudoexperiments) for all values of the nuisance parameters within that range. 
    
    The overall $p$-value for point $\thvec_i$ is
    based on the largest $p$-value within that set, $p = \max_{\nuisvec} p(\thvec_i, \nuisvec)+\beta$. This method is named after its proposers~\cite{ref:bb}. Since the nuisance parameters in the likelihood and the pseudoexperiments are moved together, this method does not have the same problem of over-coverage as the Conservative method, but it is still computationally infeasible for making confidence intervals or for a large number of nuisance parameters. Appendix~\ref{sec:bergerboos} shows the use of this method to cross-check the significance in a single hypothesis test, which is the context in which it was originally proposed.

    \item[Highland--Cousins] When generating pseudoexperiments, generate the nuisance parameters from their a priori probability distributions, $\nuisvec_i \sim P_r(\nuisvec_0)$. This method is commonly called the Highland--Cousins method after its proposers~\cite{ref:hc}. Being an explicitly hybrid Bayesian approach, its coverage properties are not guaranteed, and can be difficult to interpret in a purely frequentist framework. In the same fashion as with the a priori estimate approach, information about the nuisance parameters garnered from the experiment is here discarded, making implementation straightforward, but leading to worse performance. The Highland--Cousins method has also been shown to over-cover in circumstances where the nuisance parameter has a true fixed value but an estimated value that can vary experiment-to-experiment~\cite{ref:conrad2002,ref:conrad2004}. Since this method requires the generation of a single set of FC pseudoexperiments, it is practical to use and its coverage properties will be investigated in \sect{toy}.

    \item[A posteriori Highland--Cousins] At each point in parameter space, generate the FC pseudoexperiments with parameters drawn from the post-fit, or a posteriori, likelihood distribution derived from the observed data, $\nuisvec_i \sim P(\hat\nuisvec|\thvec_i)$. This variant has the same issue as the regular Highland--Cousins method, where the coverage is ensured for an ensemble of experiments with  nuisance parameter values drawn from the a posteriori distribution rather than considering their true values. This procedure can also be impractical to apply in frequentist analyses, which do not naturally produce these a posteriori distributions. Nonetheless, by constraining the nuisance parameter values to those most consistent with the data, the coverage for the unknown true values is likely to be more accurate. This method will be investigated in \sect{toy}.

    \item[Profile Construction] At each point in parameter space, $\thvec_{i}$, generate the FC pseudoexperiments assuming the best-fit values of the nuisance parameters, given these parameters and the observed data, $\nuisvec_i = \doublehat\nuisvec_i$, as defined in \eqn{nuisance}. This method was introduced to HEP in the PhyStat conference series~\cite{ref:phystat2003,ref:phystat2005,ref:phystat2007}, but can also be found in statistics textbooks~\cite{ref:kendall}\footnote{The examples in~\cite{ref:kendall} focused on simple cases where the probability distribution of the likelihood ratio does not depend on the nuisance parameter or where the dependence can be calculated analytically. Determining the distribution via Monte Carlo methods is only suggested by the more recent literature}. This method depends on the profiled values of nuisance parameters\footnote{The profiled nuisance parameter values are sometimes called `pull terms.'}, so it can only be applied when those values are available. For example, this method could not be applied to systematic uncertainties implemented as bin-to-bin covariance matrices, since in that case there are no explicit nuisance parameters in the likelihood\footnote{It is possible to implement hybrid versions if different parameters are treated differently. For example, in the sterile neutrino search presented in~\cite{ref:nu2022,ref:nu2022poster}, the profile construction is applied when choosing physics parameters for pseudoexperiments, but the systematic values are thrown randomly per Highland-Cousins since no pull values are available in the method used.}. This method will be investigated in \sect{toy}.
\end{description}


\section{Toy Models}\label{sec:toy}

The computational cost of many of these methods makes comparing them in situ in the full analysis prohibitive. In order to choose the best method to use in our oscillation measurements, we developed a toy model that  captures the key features of the NOvA oscillation measurement which violate Wilks' theorem. It only includes an unconstrained `physics' nuisance parameter since these were found to be the primary source of non-Wilks' behavior, and adding constrained `systematic' parameters foils the analytical calculations which make running the toy experiments computationally tractable. We developed a second toy model focused specifically on the behavior in the presence of constrained systematic uncertainties, but with a simpler linear signal and background model. This second study demonstrates how the treatment of nuisance parameters can impact the coverage, even without explicit violations of Wilks' theorem.

Source code reproducing both toy models is publicly available in~\cite{ref:code}.

\subsection{NOvA-like Toy Model}

The toy consists of the measurement of a single number -- the number of events observed. We take the expected number to be given by
\begin{linenomath*}
\begin{equation}
    \nexp = A - B\sin\delta \pm C,
\end{equation}
\end{linenomath*}
where $A$, $B$, and $C$ are fixed constants and the expectation, $\nexp$ depends on a 2 unknown parameters: a continuous, cyclic parameter, $\delta$, and a binary parameter corresponding to a positive or negative sign for the $C$ term.

\begin{figure}
    \centering
    \includegraphics[width=.55\textwidth]{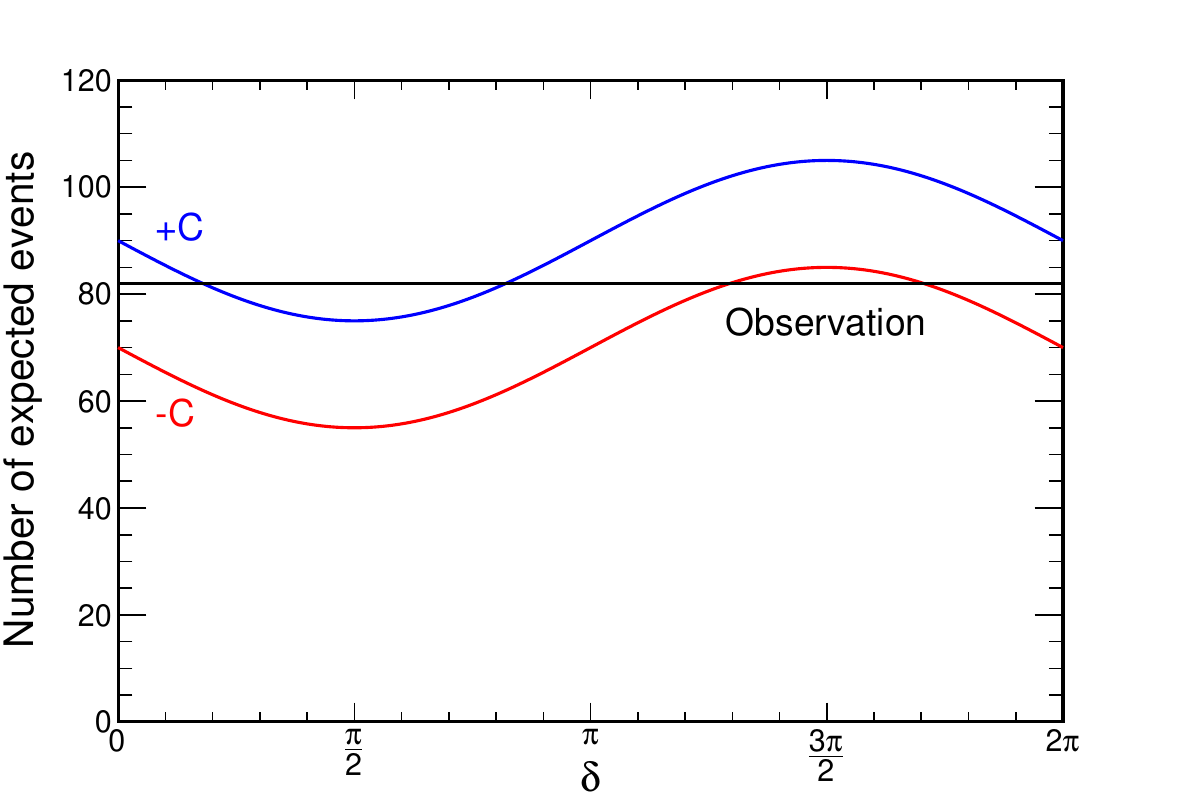}
    \caption{The number of events expected in the toy model as a function of the continuous $\delta$ parameter ($x$-axis) and sign of $C$ term (positive sign blue, negative sign red). A hypothetical observation of a particular number of events is shown in black. }
    \label{fig:toy_nexp}
\end{figure}

We choose values for the constants:
\begin{eqnarray*}
  A &=& 80,\\
  B &=& 15,\\
  C &=& 10,
\end{eqnarray*}
so that the toy model has event counts similar to current rates from the NOvA experiment~\cite{ref:nova3f2021}. \fig{toy_nexp} illustrates this function, along with a hypothetical measurement that we would want to interpret. The experiment consists of making a single measurement of the number of events observed, \nobs, comparing to the expected number of events \nexp, and using that to generate confidence regions in $\delta$ or determine the sign of the $C$ term.

Constraining ourselves for the moment to the case where the sign of $C$ is already known (we have external information telling us for certain which sign to pick) one derives a confidence interval by first finding the value \hatdelta that provides the best match to the observed data (the best fit given \nobs), and then computing:
\begin{linenomath*}
\begin{equation}
    \LR(\delta) = \LL(\delta) - \LL(\hatdelta)
\end{equation}
\end{linenomath*}
for each value of $\delta$ under consideration.

For the purposes of keeping this toy minimal, and to avoid discontinuities arising from discrete event counts\footnote{Typical physics analyses have many bins and continuous parameters. But the first NOvA electron neutrino appearance data, with only a handful of events in each bin, caused discontinuities to appear. An example of this type of discontinuity caused by integer event counts can be seen in Fig. 4 of~\cite{ref:first_nue}.}, we will assume \nobs is normally distributed with mean \nexp and standard deviation $\sqrt{\nexp}$, and thus:
\begin{linenomath*}
\begin{equation}
    \LL(\delta)=\frac{\Big(\nexp(\delta)-\nobs\Big)^2}{\nexp(\delta)} .
\end{equation}
\end{linenomath*}
To determine confidence intervals, one then compares $\LR(\delta)$ to \crit and accepts all values of $\delta$ having a lower \LR. According to Wilks' theorem, $\LR \sim \chisq_{k=1}$, and one should therefore use $\crit=1$ to achieve 68.27\% coverage.

\begin{figure}
    \centering
    \begin{minipage}[t]{.49\linewidth}
    \includegraphics[width=\linewidth]{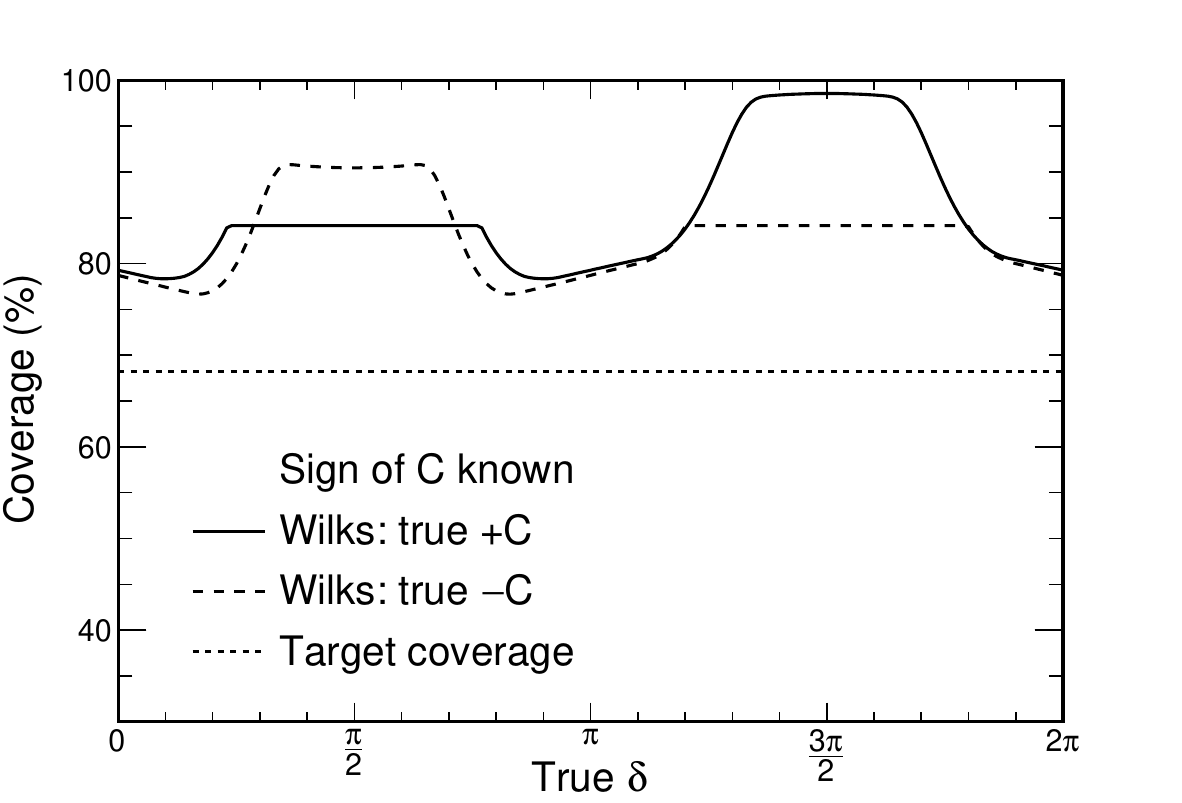}
    \caption{Coverage for the toy experiments using Wilks' theorem in the case where the true sign of $C$ is positive and this fact is known to the fitter (solid) and likewise true $-C$ known to the fitter (dashed). The short-dashed line indicates the desired coverage. Since there are no nuisance parameters, all other discussed techniques are equivalent to Feldman-Cousins. Since they would all perfectly match the target coverage, they are not shown in this figure.} 
    \label{fig:toy_known_hier_cov}
    \end{minipage}
    \hfill
    \begin{minipage}[t]{.49\linewidth}
    \includegraphics[width=\linewidth]{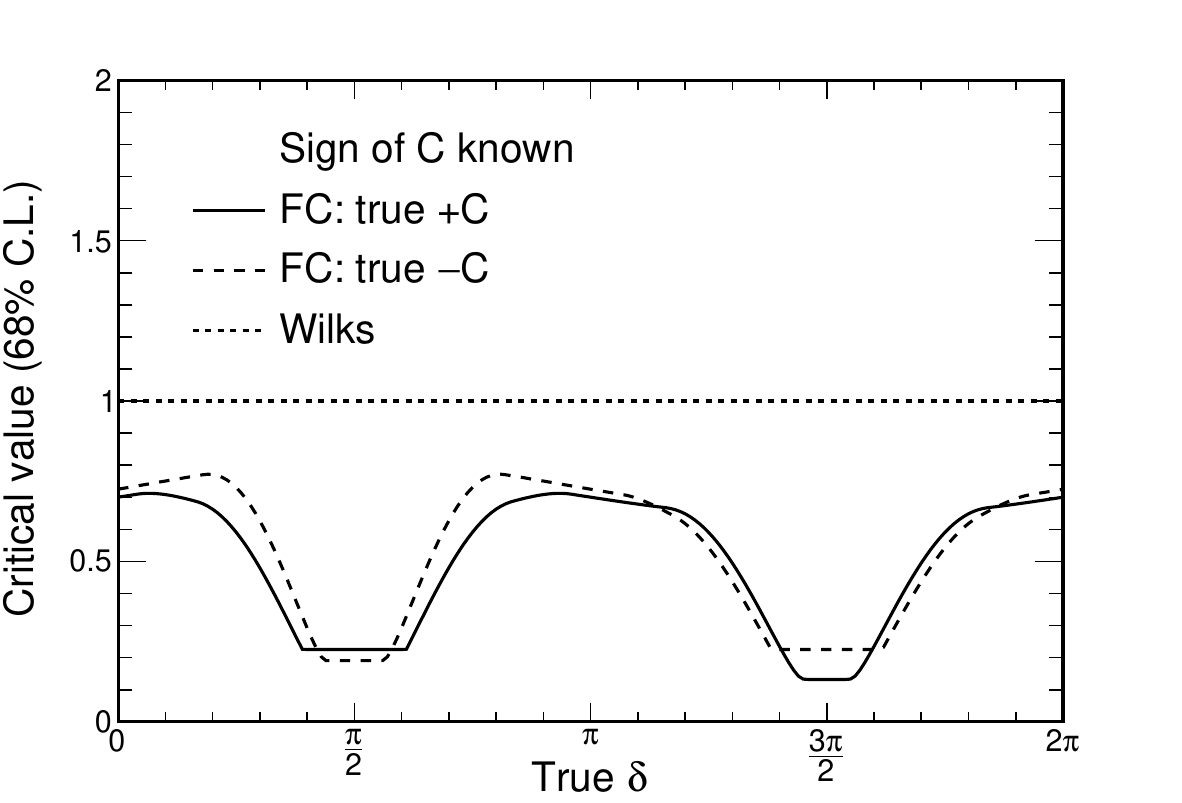}
    \caption{Critical values evaluated for the toy experiments using the Feldman--Cousins procedure in the case where the true sign of $C$ is positive and this fact is known to the fitter (solid) and likewise true $-C$ known to the fitter (dashed). The critical value shows substantial deviations from the expectation of Wilks' theorem (short-dashed) in those regions where the Wilks' critical value most over-covered. 
    }
    \label{fig:toy_known_hier_crit}
    \end{minipage}
\end{figure}

This Wilks' procedure over-covers significantly, even when the sign of $C$ is known in advance. 
The over-coverage comes from two sources. The first is a degeneracy affecting all true values of $\delta$: any observation, \nexp, within the expected model range $A-B+C< \nexp <A+B+C$ for positive $C$, is consistent with two different values of $\delta$ 
due to the periodic nature of the \nexp function. 
The second occurs in cases where, through random chance, the observed data might be outside the model range. When that occurs, the \nexp is not perfectly compatible with any $\delta$ and the minimum $\LL(\hatdelta)$ will always be found
at the extreme of the function range, making $\LL(\hatdelta)$ larger than it would be without constraints, and causing a larger region of the $\delta$ space to have a value of \LR below 1. This \term{physical boundary} effect is expected to be largest when the true value of $\delta$ is near $\pi/2$ or $3\pi/2$, where such a fluctuation is expected to occur 50\% of the time. \fig{toy_known_hier_cov} shows this over-coverage vs.\ the true value of $\delta$. We evaluate coverage by generating a series of statistically fluctuated toy experiments at each true value of $\delta$, determining the best fit and confidence interval that would be obtained for each, using $c_{68\%}=1$, and counting the fraction of these toy experiments in which the true $\delta$ value is included in the confidence interval.

In this circumstance where the sign of $C$ is known, the Feldman--Cousins procedure can be followed to produce perfect coverage for any value of $\delta$. \fig{toy_known_hier_crit} shows how the critical value, $c_{68\%}$, varies as a function of $\delta$, with substantially lower critical values in the regions nearest the physical boundary to account for the effect described above. Using these critical values to evaluate the coverage of an independent set of mock experiments yields ideal coverage, as would be expected in this case since the FC pseudoexperiments were generated in exactly the same way.

\begin{figure}
    \centering
    \includegraphics[width=.5\linewidth]{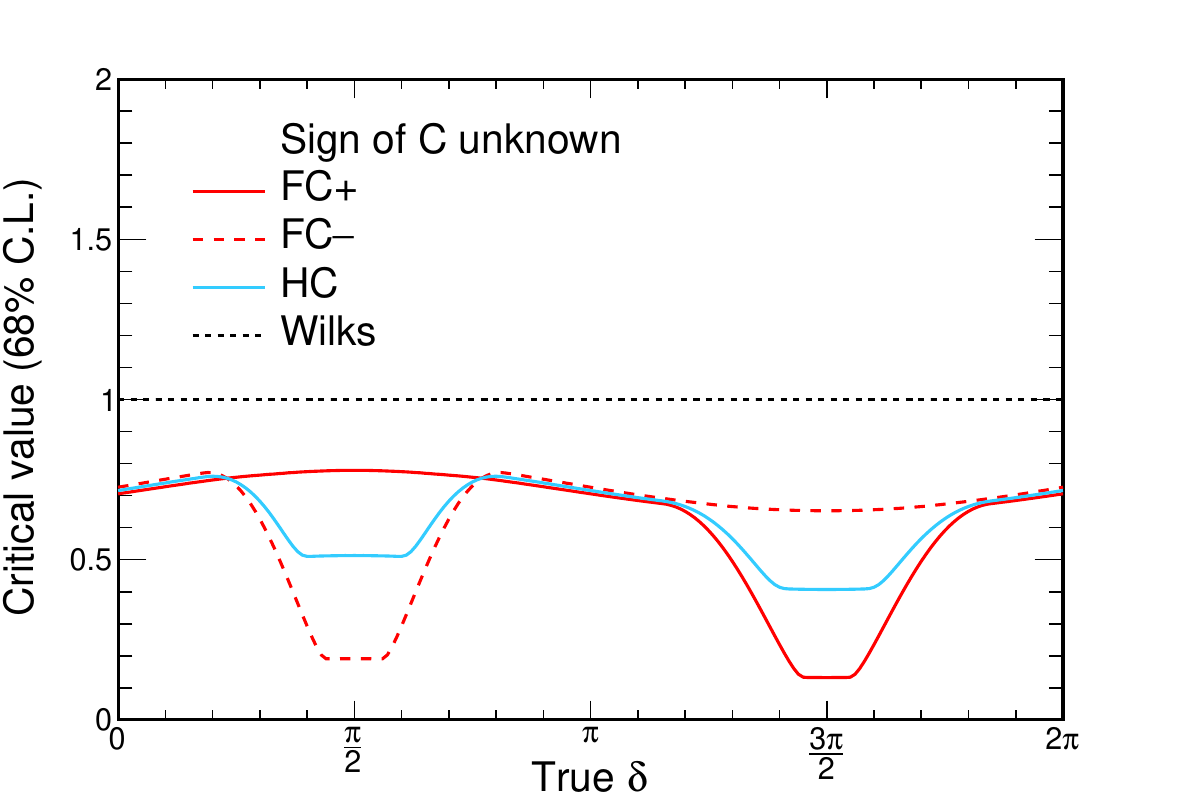}
    \caption{Critical values for the 68\% C.L. from Wilks' theorem (the horizontal black line at 1), the Feldman--Cousins procedure (red) and Highland--Cousins (light blue), where the true sign of C is positive. The Feldman--Cousins critical values are shown for two cases -- generating the FC pseudoexperiments assuming positive $C$ (solid) and assuming negative $C$ (dashed). In our toy model, the Highland--Cousins procedure consists of generating the FC pseudoexperiments with an equal mixture of the two signs, and the blue curve splits the difference between the red curves as expected. The profile construction cannot be displayed on this plot; it amounts to choosing one or other of the Feldman--Cousins curves at each value of $\delta$ depending on the observed data.}
    \label{fig:comp_crit}
\end{figure}

In the full experiment, we do not know the true sign of $C$. One common approach is to present the results for both possible choices of the binary parameter. However, if the results for the parameter $\delta$ are desired irrespective of that choice, another common frequentist procedure is to profile over the sign of the parameter,
\begin{linenomath*}
\begin{equation}
    \LL(\delta) = \min\Big(\LL^+(\delta), \LL^-(\delta)\Big),
\end{equation}
\end{linenomath*}
where $\LL^+$ is evaluated using the values of \nexp based on the positive sign for $C$, and similarly for $\LL^-$. We can replicate this procedure in the fits performed on the FC pseudoexperiments, but we are still left with the question of how to generate the FC pseudoexperiments. We will obtain different critical values if we generate all the FC pseudoexperiments with positive vs.~negative sign, as shown by the solid and dashed lines in \fig{comp_crit}, because the boundaries on possible values of \nexp are now wider ($A-B-C< \nexp <A+B+C$), and FC pseudoexperiments generated assuming a particular sign will only run up against one boundary. The previous example where the sign was known (\fig{toy_known_hier_crit}) showed large downward deviations in the critical value at both $\pi/2$ and $3\pi/2$ since both were boundaries on \nexp, but now there is only a large deviation at $3\pi/2$ for the positive sign, where it runs into the high-side boundary on \nexp, and at $\pi/2$ for the negative sign where it runs into the low-side boundary.
In the intermediate regions around 0, $\pi$, and $2\pi$, where the event counts in the pseudoexperiments will typically be far from the overall upper and lower limits no matter which sign we assume when generating them, the critical values closely follow each other. 

\begin{figure}
    \centering
    \begin{minipage}[t]{.49\textwidth}
    \includegraphics[width=\textwidth]{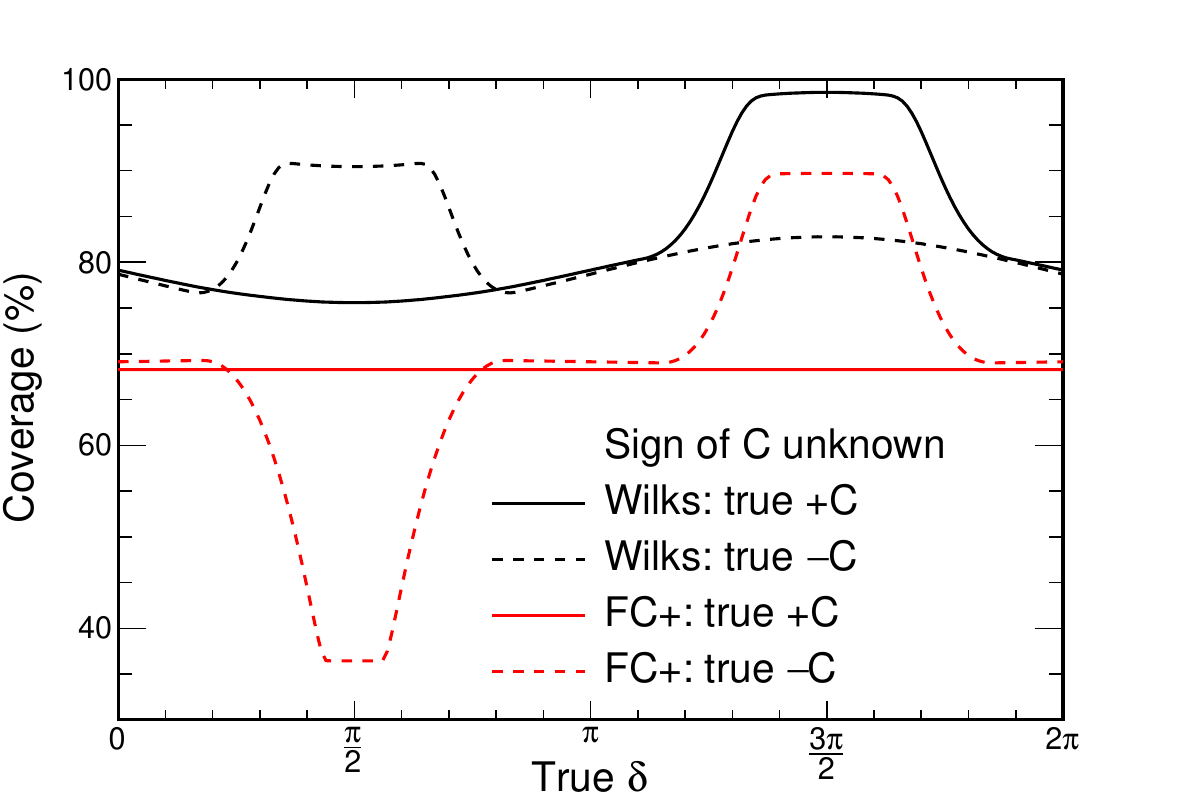}
    \caption{The coverage obtained for our toy experiments using critical values from Wilks' theorem (black) and the Feldman--Cousins procedure, which here assumes a positive sign for $C$ for the FC pseudoexperiments (red). Coverage is shown vs.~true $\delta$ and true sign (solid/dashed for positive/negative). The true sign is \emph{not} known at fit time and is profiled over. The Wilks' theorem critical values lead to substantial over-coverage in all cases. Since the FC pseudoexperiments have been generated assuming positive sign, the procedure produces exactly the target coverage of 68\% for toy experiments with true positive sign, but for true negative sign the coverage properties are particularly poor.}
    \label{fig:comp_cov_wilksfc}
    \end{minipage}
    \hfill
    \begin{minipage}[t]{.49\textwidth}
    \includegraphics[width=\textwidth]{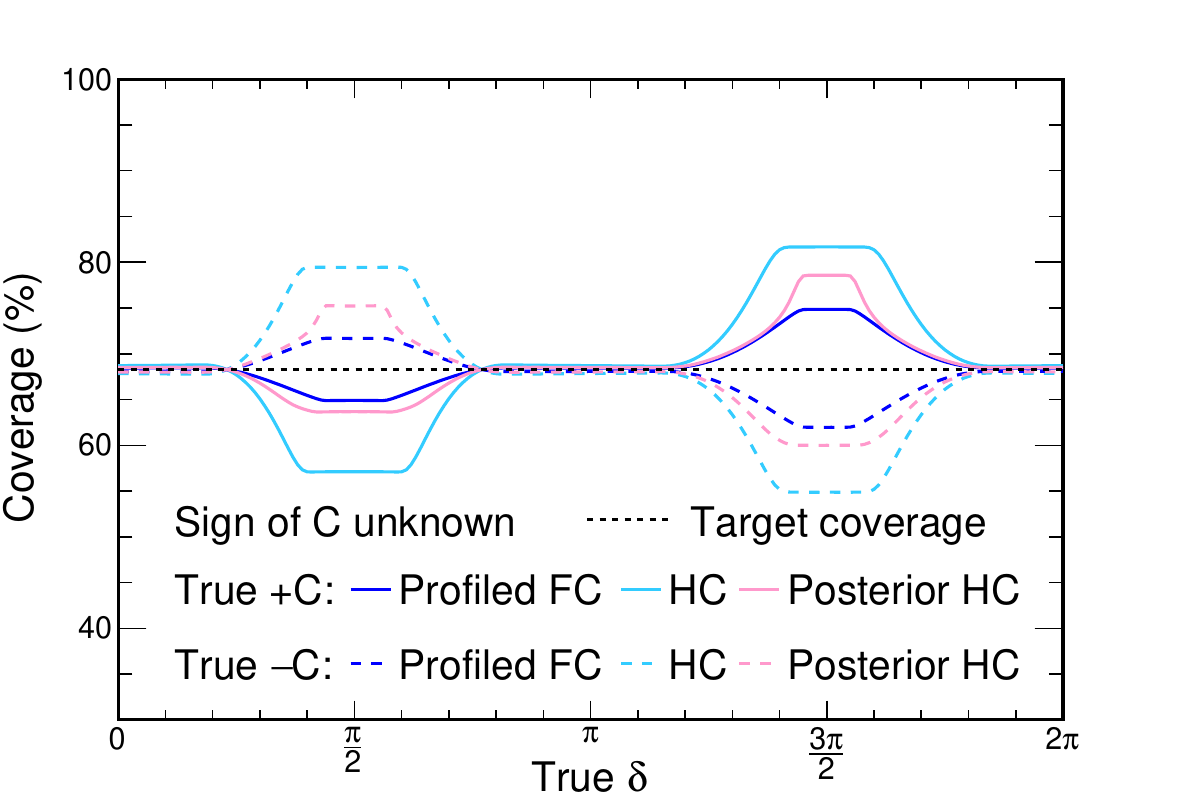}
    \caption{The coverage obtained using critical values using the Highland--Cousins procedure (light blue) and the profile construction (dark blue) for our toy model, where the true sign of $C$ is unknown at fit time, and profiled over, evaluated for true positive sign (solid) and true negative sign (dashed). In both cases the coverage averaged over $\delta$ and sign is correct, but the profiling procedure exhibits substantially smaller deviations from correct coverage where these occur. Also shown is the a posteriori Highland-Cousins method (here labeled \term{Posterior HC} and drawn in pink) which can be considered as an intermediate option between Highland--Cousins and the profile construction, and yields an intermediate performance.}
    \label{fig:comp_cov_hcprof}
    \end{minipage}
\end{figure}

The consequences of this behavior for the coverage of confidence intervals are shown in \fig{comp_cov_wilksfc}, which compares the coverage vs.~true values of $\delta$ and sign of $C$ (solid/dashed for positive/negative) from Wilks' theorem (black) and from the Feldman--Cousins procedure where we arbitrarily choose to generate FC pseudoexperiments assuming the positive sign. As in \fig{toy_known_hier_cov}, Wilks' theorem shows over-coverage everywhere, but it is substantially worse when the true values lie near the boundaries on \nexp ($+C$, $\delta = 3\pi/2$ or $-C$, $\delta=\pi/2$). The Feldman--Cousins method yields ideal coverage in the $+C$ case, but large deviations in the case of true $-C$, where the FC pseudoexperiments have incorrectly encountered a physical boundary (at $3\pi/2$) or missed one (at $\pi/2$). The results for experiments generated assuming negative sign show the same qualitative behaviour, but with the roles of $\delta=\frac{\pi}{2}$ and $\delta=\frac{3\pi}{2}$ reversed.

For the present toy experiment, the Highland--Cousins procedure consists of splitting the difference by generating the FC pseudoexperiments equally from each sign (assuming a 50:50 prior expectation). This has the predictable effect of yielding critical values intermediate between the FC expectations from the two signs (light blue line in \fig{comp_crit}) and coverage (light blue lines in \fig{comp_cov_hcprof}) intermediate between the \term{right} and \term{wrong} FC coverage (red lines, solid and dashed respectively, in \fig{comp_cov_wilksfc}). This is certainly an improvement from the FC$^+$ (or FC$^-$) case -- the \term{average} coverage is correct, and there is no longer a large difference in behaviour depending on the true sign.

The profile construction achieves better results than any of these methods by using information from the observed data itself. If we observe a large number of events, say $\gtrsim85$, we know it is more likely that the critical value evaluated under the $+C$ hypothesis will provide the right coverage, and similarly a small number of observed events, $\lesssim70$, suggests the $-C$ hypothesis is more likely to provide correct coverage. If we observe an intermediate number of events (values close to 80), then we have gained no information about the true sign of $C$, but in that case the critical values are very similar either way.

In this case, for each toy experiment contributing to the coverage evaluation, for each value of $\delta$ whose membership in the confidence interval we need to determine, we evaluate which sign gives the best match (lowest $\LL$) to the data, and generate the FC pseudoexperiments from which the critical value will be derived assuming that sign. For a continuous nuisance parameter, we would generate experiments assuming the best-fit value.

The blue lines in  \fig{comp_cov_hcprof} show the coverage obtained by this procedure. Deviations still occur in the regions where the two critical values differ, but the magnitude is substantially reduced compared to Highland--Cousins. The remaining mis-coverage is due to those cases where a statistical fluctuation produces a number of events more compatible with positive sign, despite the true sign being negative, or vice versa.

The Posterior Highland--Cousins approach -- generating the FC pseudoexperiments distributed between the two signs based on the posterior distribution -- represents an intermediate point between Highland--Cousins (generating pseudoexperiments equally from the two signs) and our profiling method (generating pseudoexperiments from the best-fit sign). Unsurprisingly, for these toy experiments it yields intermediate coverage properties -- better than Highland--Cousins but not as good as the profile construction.

\subsection{Toy Model with Constrained Systematic Uncertainties}

A second toy model was developed to study the coverage properties of the profile construction, Highland-Cousins, and Wilks' theorem in the presence of constrained nuisance parameters, a common method for implementing systematic uncertainties. In order to make the calculations tractable, the model itself is simpler than the NOvA-like case above. Here, we take the expected number of observed events, $N_{\textrm{exp}}$, to be:
\begin{equation}
    N_{\textrm{exp}} = S + B
\end{equation}
where $S$ refers to signal and $B$ refers to background, where $B$ has been externally constrained to a value $B_0$ with uncertainty \sigsy. As above, we will assume that the data is normally distributed (and we use large enough numbers in the concrete examples below for this to be a good approximation), so the log-likelihood function is:
\begin{eqnarray}
    \LL(S, B| \Ndata) &=& \displaystyle\frac{(\Ndata - S - B)^2}{N} + \frac{(B - B_0)^2}{\sigsy^2} \\
    \LL(S | \Ndata) &=& \displaystyle\min_B \LL(S,B|\Ndata) \\
                      &=& \LL(S, \hat B(S|\Ndata) | \Ndata)
\end{eqnarray}
where the second \LL function has profiled over the nuisance parameter, $B$. In this simple example, $\hat B$ can be calculated analytically given the other parameters defined above and an observed \Ndata by finding the root of the derivative of \LL with respect to $B$:
\begin{equation}
    \hat B(S|\Ndata) = \frac{\Ndata B_0 + (\Ndata - S)\sigsy^2}{N + \sigsy^2},
\end{equation}
and the maximum likelihood estimate (or best fit point) for the signal, $\hat S$ will be at the point where both terms in the \LL equal 0:
\begin{equation}
    \hat S = \Ndata - B_0.
\end{equation}

The coverage accuracy was estimated by testing every possible integer value of \Ndata between $\pm 5.5\sigst$ on $\Ndata = S + B$, weighted by the likelihood of having drawn that particular value of \Ndata from a normal distribution centered on $S+B$\footnote{Equivalent results, but with more noise, are obtained by randomly drawing \Ndata values from a Poisson distribution with rate $\lambda = S+B$.}. The coverage accuracy for a $p$-value is defined as:
\begin{equation}
    \textrm{Coverage Accuracy} = \frac{C - (1-p) }{p},
\end{equation}
where $C$ is the observed frequency at which the true value $S$ is included in the confidence intervals in the weighted toy experiments; perfect coverage is achieved when $C = 1 - p$. $C$ is calculated with:
\begin{equation}
    C = \frac{\sum_{i} w_i \Theta(p_i - p)}{\sum_{i} w_i},
\end{equation}
where $i$ steps through the possible values of \Ndata, $\Theta()$ is the Heaviside step function that is 1 if its argument is 0 or greater and 0 otherwise, $p_i$ is the $p$-value calculated for the true value of $S$ for toy experiment $i$ for a given method, and $w_i$ is the weight for that experiment. The weight is defined as:
\begin{equation}
    w_i = \frac{\mathcal{N}\left(N_i \mid S+B, \sqrt{S+B}\right)}{\mathcal{N}\left(N_0 \mid S+B, \sqrt{S+B}\right)}
\end{equation}
where $\mathcal{N}$ is the PDF of the normal distribution, $N_i$ is the value of \Ndata and the denominator is the value of the smallest (i.e. least probable) \Ndata. This weight function assigns a weight of 1 to $N_0$ and weights up other experiments by how much more frequently they should be sampled relative to $N_0$.

\begin{figure}
    \centering
    \includegraphics[width=\textwidth]{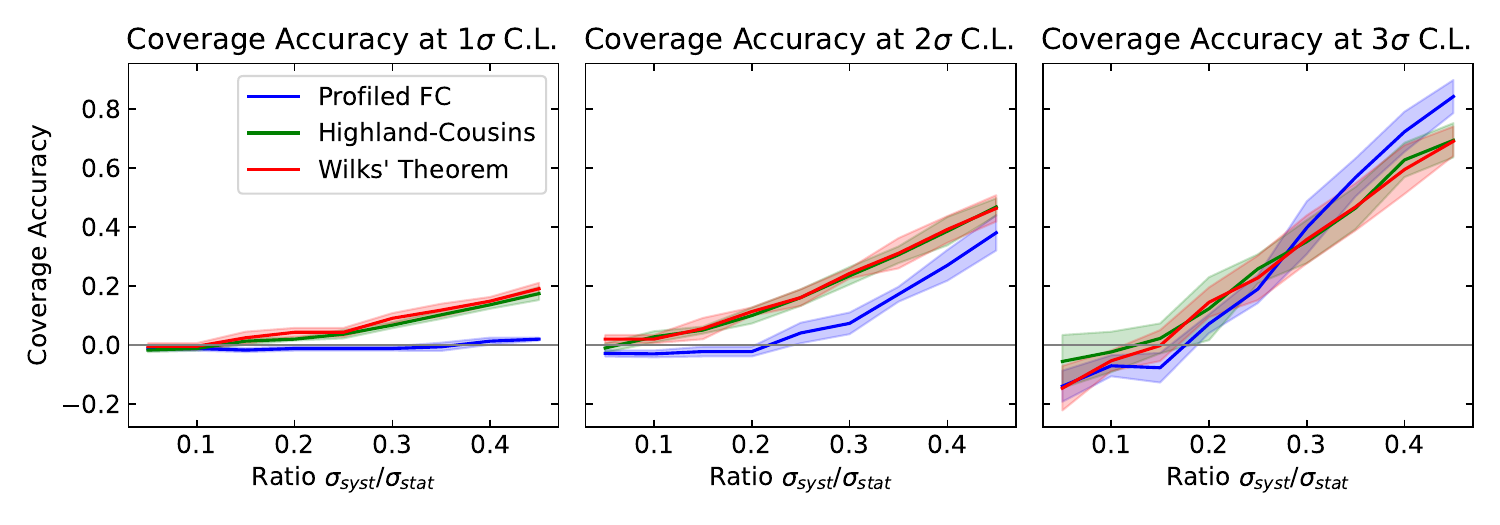}
    \caption{These plots show the accuracy of coverage of 1-, 2-, and 3-$\sigma$ confidence intervals for the the profile construction (blue), Highland-Cousins (green), and Wilks' Theorem (red) methods, plotted vs. the relative size of the systematic and statistical error on the measured parameter, $S$. A range of different signal:background balances and systematic uncertainty sizes were tested, and the mean and standard deviation across those different tests are plotted here, showing that the ratio on the x-axis is the key independent variable.}
    \label{fig:acc_syst_bkgd}
\end{figure}

\begin{figure}
    \centering
    \includegraphics[width=\textwidth]{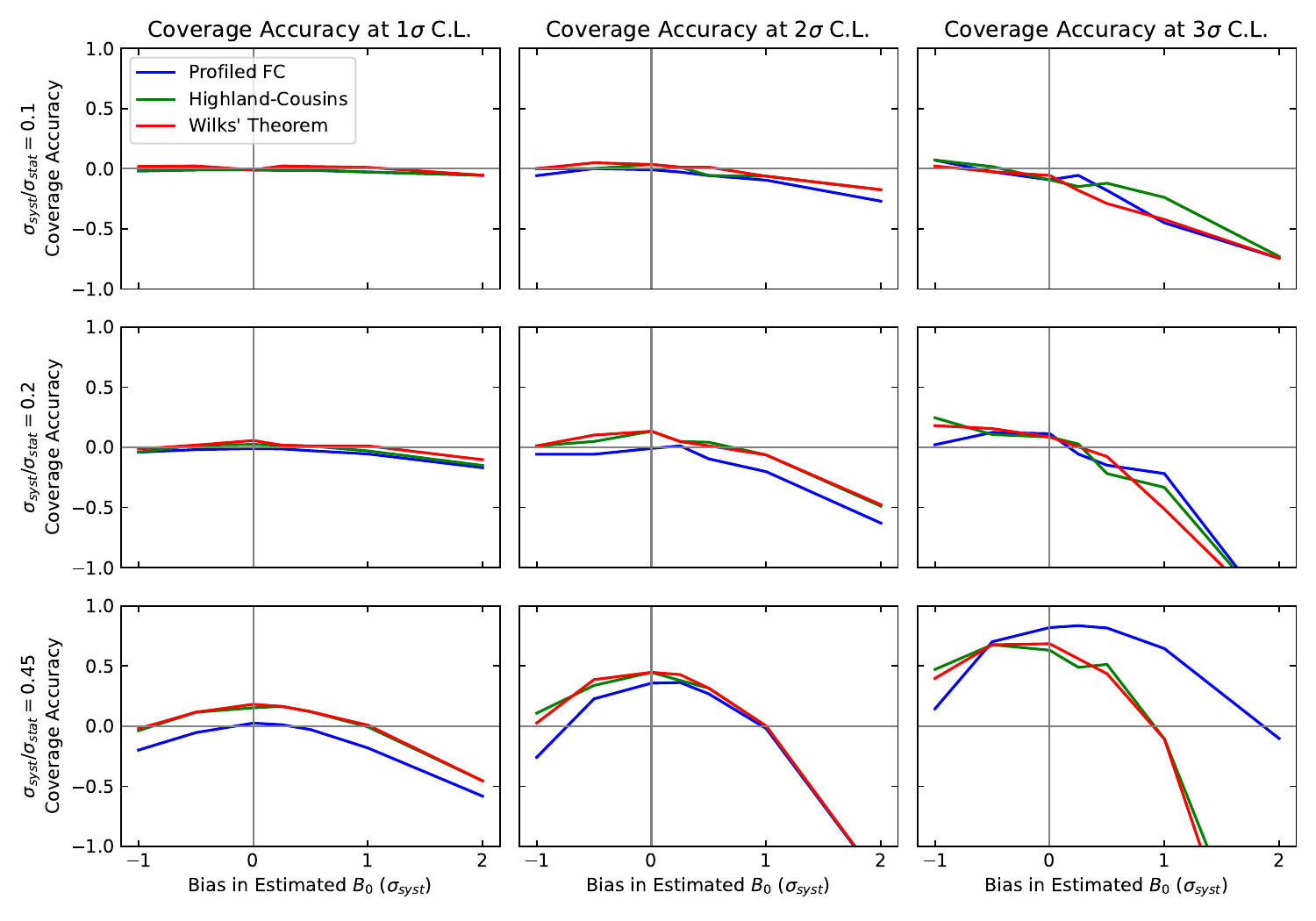}
    \caption{These plots show the accuracy of coverage of 1-, 2-, and 3-$\sigma$ confidence intervals for the the profile construction (blue), Highland-Cousins (green), and Wilks' Theorem (red) methods, plotted vs. the bias in the estimate of the background parameter $B_0$ in units of \sigsy, for 3 different values of the relative size of the systematic and statistical error on the measured parameter, $S$. For biases below 1\sigsy and relative systematic uncertainties of $20\%$ or below, all the methods give reasonably accurate coverage. As the systematic uncertainties and biases increase, all the methods have worse coverage accuracy, with none performing obviously better in these challenging scenarios.}
    \label{fig:acc_syst_bias}
\end{figure}

For this toy model, it is straightforward to test the coverage of the profile construction, Wilks' theorem, and Highland-Cousins methods given some specific values for the parameters above\footnote{For the Highland-Cousins and profile construction methods, 100,000 pseudo experiments were generated. For HC, this only needs to be done once for each set of parameters, while for the profile construction, the pseudoexperiments are thrown for each possible value of \Ndata.}. After testing a variety of possible choices, we determined that the key parameters defining the coverage behavior were the size of the systematic uncertainty, \sigsy relative to the size of the statistical error on $S$,
\begin{eqnarray}
    r &=& \sigsy / \sigst \\
      &=& \sigsy / \sqrt{S + 2B}
\end{eqnarray}
and the bias in the external estimate of the nuisance parameter $B_0$, relative to \sigsy:
\begin{equation}
    b = \frac{B_0 - B_{\textrm{true}}}{\sigsy}.
\end{equation}
With those ratios held fixed, changing the specific number values of $S$ and $B$ did not affect the coverage accuracy, as long as the numbers chosen were large enough to avoid significant deviations between the normal and Poisson distributions. For all experiments shown here, an $S = 350$ was used. \fig{acc_syst_bkgd} shows the results for $B = \{50, \allowbreak 150, \allowbreak 250, \allowbreak 350\}$ and $\sigsy = \{5\%, \allowbreak 10\%, \allowbreak 15\%, \allowbreak 20\%, \allowbreak 25\%, \allowbreak 30\%, \allowbreak 35\%, \allowbreak 40\%, \allowbreak 45\%\}$, where the lines represent the mean and the shaded region shows the standard deviation across these different combinations of values. The narrow size of the standard deviation shows that the ratio $r$ above is the key independent variable driving behavior. 

\fig{acc_syst_bkgd} shows that coverage performance is better at low significances and where systematic uncertainties are small relative to statistical uncertainties across all methods. 
For systematic uncertainties below $20\%$ of the statistical error, all the methods give reasonably good coverage accuracy.
At the $1\sigma$ and $2\sigma$ significance levels, the profile construction gives the most accurate coverage among the 3 methods tested across a range of systematic uncertainty sizes. At $3\sigma$ significance, all 3 methods have equivalent coverage performance. 

Based on these results, the study of the bias, $b$, in the systematic estimate shown in \fig{acc_syst_bias} was only performed at $B = 150$ and the $\sigsy = {10\%, 20\%, 45}\%$. The three methods show very similar behavior when systematic uncertainties are small relative to statistical errors (first two rows), with all methods under-covering for very large positive biases in the estimated background, $B_0$. When systematics are large (third row) and biases are large, accurate coverage becomes similarly challenging for all methods.

\FloatBarrier


\section{Implementation in the NOvA Analysis}\label{sec:nova}

The primary goal of a neutrino oscillation experiment like NOvA is to measure the parameters which govern neutrino oscillations, namely the mixing angles and phase from the PMNS mixing matrix as well as the differences between the neutrino masses~\cite{ref:nova3f2021}. Additionally, certain \term{binary} questions can be addressed: whether the ordering of the neutrino masses is \term{normal} or \term{inverted,} i.e., whether $m_3$ is larger or smaller than $m_1$, or whether the mixing angle $\theta_{23}$ is larger or smaller than $45^\circ$, referred to as the upper and lower `octant' of that angle. The parameters of the toy experiments in the previous section correspond to some of these parameters: $\delta$ plays the role of the PMNS phase, \dcp, while the sign of $C$ could refer to either of the mass ordering or the octant.

These parameters, as described above, cannot be observed directly. Instead, the experiment uses a beam of muon (anti)neutrinos~\cite{ref:numi} and measures the rate of disappearance of muon (anti)neutrinos and the rate of appearance of electron (anti)neutrinos as a function of their estimated energy. Since the parameters of interest govern these disappearance and appearance rates, they can be estimated from the observed energy spectra via Maximum Likelihood Estimation~\cite{ref:pdg}. The confidence intervals describing the uncertainty on these parameters are then determined using the methods described here.

After some concrete illustrations of how Wilks' conditions are not satisfied, this section describes some key technical details in the implementation of the profile construction in the NOvA oscillation analysis. Substantially more details on the optimization of this method to run on High Performance Computing platforms will be available in an upcoming paper.

\subsection{Violations of Wilks' theorem assumptions in NOvA's neutrino oscillation analysis}\label{sec:nova_wilks}
Feldman and Cousins first introduced the FC method in the context of a neutrino experiment~\cite{ref:fcnomad} where the conditions for Wilks' theorem, described in Section~\ref{sec:orig_FC} were not met. The NOvA 3-flavor oscillation analysis violates these three conditions as follows:

(1) Effective boundaries: Many of the parameters of the oscillation model have effective boundaries of some kind. One example can be seen with the 2-flavor approximation of the survival probability for neutrino flavor $\nu_\alpha$:
\begin{linenomath*}
\begin{equation}
    P(\nu_\alpha \rightarrow \nu_\alpha) = 1 - \sin^2(2\theta)\sin^2\!\left(\frac{\dmsq{} L}{4E}\right),
\end{equation}
\end{linenomath*}
where $L$ is the constant distance, $E$ is the neutrino energy, and \dmsq{} and $\theta$ are the independent parameters being measured. While the angle $\theta$ is unconstrained, the impact it has on the observable (the survival probability) is constrained by unitarity: if $\theta = \pi/4$, either increasing or decreasing $\theta$ will lead to a reduction in the oscillation probability. This effect can be see on the right side of Figure~\ref{fig:sig_plots}. Similarly, the $\mathcal{CP}$-violating phase \dcp is cyclic and not well constrained, so it also easily runs up against effective \term{boundaries} in its possible impact.

(2) Nested hypotheses: The nested hypothesis assumption is not violated for all measurements, but it is clearly violated for binary questions. When there are only 2 possible disjoint outcomes (e.g. mass ordering is normal vs. inverted or upper vs. lower octant of $\theta_{23}$), whichever is chosen as the null cannot be a special case of the alternate. In practice, confidence intervals showing both (or all 4) choices are presented where possible, but profiling over the octant is necessary when determining the mass hierarchy significance (discussed in detail in Section~\ref{sec:hypothesis}) and is used for some other significance plots as well.

The procedure followed by NOvA is presented next.

\subsection{Fitting the data}

NOvA measures the energy spectra of disappearing muon (anti)neutrinos and appearing electron (anti)neutrinos in order to constrain parameters of the neutrino oscillation model: the mixing angle $\theta_{23}$, the mass splitting \dmsq{32}, in particular its sign, equivalent to determining the neutrino mass ordering, and the CP--violating phase \dcp. The candidate neutrino interactions are divided into different categories (based on energy resolution and particle identification criteria) to optimize the measurement's sensitivity. The relative compatibility between model predictions given sets of parameter values and some data is quantified with a likelihood function $\mathcal{L}$. The best fit is found by maximizing $\mathcal{L}$, or minimizing $\ell = \logl$.
Since the data is structured as a histogram (meaning a set of counts of independent events), the likelihood function for Poisson--distributed data~\cite{ref:pdg} is used\footnote{Or more accurately $\ell = \logll$, where $\mathcal{L}_0$ is the likelihood when $o_{i}=e_{i}$ }:
\begin{linenomath*}
\begin{equation}
\LL_{stat} = 2\sum_i\left( e_i(\thvec) - o_i + o_i\ln\frac{o_i}{e_i(\thvec)}\right),
\label{eqn:poissonll}
\end{equation}
\end{linenomath*}
where $e_i(\thvec)$ is the expected number of events in bin $i$ given parameter values \thvec, and $o_i$ is the observed number of events in that same bin. The $e_i(\thvec)$'s are calculated by extrapolating the muon (anti)neutrino energy spectrum measured in NOvA's near detector to its far detector assuming a set of neutrino oscillation parameters, taking into account known differences in flux and acceptance between the detectors. In addition to the oscillation parameters, around 50 systematic uncertainties are included in the fit as nuisance parameters, with penalty terms added to the likelihood in \eqn{poissonll}:
\begin{linenomath*}
\begin{equation}
    \LL = \LL_{stat} + \sum_k \frac{\phi_k^2}{\sigma_k^2},
\end{equation}
\end{linenomath*}
where $\sigma_k$ is the prior uncertainty on the $k^{\textrm{th}}$ nuisance parameter $\phi_k$. The sources of uncertainty vary from parameter to parameter. For example, some uncertainties are based on the uncertainties quoted by external measurements, some are based on the level of agreement between data and simulation within the experiment, and some are based on comparisons between alternative theoretical models. The values of \sinsq{23}, \dmsq{32}, and \dcp which minimize \LL (i.e., the Maximum Likelihood Estimate or best fit point) are found using the Minuit2 minimizer~\cite{ref:minuit}. This best fit point is the basis from which the confidence intervals and significances, the main topic of this paper and main results of the oscillation analysis, are constructed. 

\subsection{Building 1-dimensional and 2-dimensional confidence intervals}
To build 1-dimensional or 2-dimensional maps of the significance, we need to sample the oscillation parameter space finely enough to catch possible local features, while also being limited by the computational costs the Profiled Feldman--Cousins approach entails. In practice, this means that the significance is evaluated at  60 points evenly distributed across the range of parameter values when building 1-dimensional significance maps. These one-dimensional plots can be constructed with the parameters constrained in one mass ordering, one \thetamix octant\footnote{$\theta_{23}<45^{\circ}$ is commonly referred to as the lower octant, while $\theta_{23}>45^{\circ}$ is the upper octant.}, or a combination of both. In two dimensions, we report confidence intervals (i.e., contours) for \sinsq{23} vs.~\dcp (estimated in a 30$\times$30 grid) and \dmsq{32} vs.~\sinsq{23} (in a 20$\times$20 grid), for both orderings.

As explained earlier, we chose to profile the nuisance parameters. The first step is therefore to fit the data with the parameters of interest fixed at each grid point, $\thvec_i$, and find $\doublehat\nuisvec_i$, the set of nuisance parameters minimizing \LL~per \eqn{nuisance}. This process can be conveniently run on standard distributed computing resources and serves as an input to the more computationally intensive generation and fitting of millions of Feldman--Cousins pseudoexperiments in a High Performance Computing environment.
From that first step, we can already obtain maps of the significance under Wilks' theorem, which provides a good first approximation of the final significance. The Feldman--Cousins procedure then modifies those maps, increasing or decreasing the significance depending on the distribution of the underlying test statistic, which is why this procedure can be considered a correction. We can also take advantage of those approximated significances to estimate the number of FC pseudoexperiments that need to be generated at each point of the parameter space, $\thvec_{i}$, to reach a desired statistical accuracy when measuring the $p$-values from the empirical \LR distributions. Working backwards from the formulation of the binomial uncertainty, the number of pseudoexperiments, $N_{PSE}$, required to reach an uncertainty $u$ can be expressed as:
\begin{equation}
    N_{PSE} = \left[ u^2 \left( Q \left(\frac{n_{dof}}{2}, \frac{\lambda_{Wilks}}{2} \right) - 1 \right) \left( Q \left( \frac{n_{dof}}{2}, \frac{\lambda_{Wilks}}{2} \right) \right) \right]^{-1},
    \label{eqn:npse}
\end{equation}
where $\lambda_{Wilks}$ is estimated from the data under Wilks' conditions, $n_{dof}$ is the number of degrees of freedom, and $Q$ is the regularized incomplete gamma function. In practice, we require a relative uncertainty on the p-value no greater than 5\%, which translates to a few thousand pseudoexperiments. This uncertainty is chosen to be negligible compared to the other measurement's uncertainties. The number of pseudoexperiments is capped at 5,000 for a single point of the parameter space because of computational constraints, which means 3-sigma regions could be described with a lesser accuracy, albeit still not constituting the dominating source of uncertainty.
For each $\thvec_{i}$, the FC pseudoexperiments are constructed by generating Poisson--fluctuated neutrino energy spectra from the predictions made at ($\thvec_i, \doublehat{\nuisvec}_i$) determined above. For each FC pseudoexperiment, $j$, generated at point $i$, a likelihood ratio is estimated:
\begin{linenomath*}
\begin{align}
\begin{split}
    \LR_{ij} &= \LL_\mathrm{constrained} -\LL_\mathrm{unconstrained}\\
    &=\LL(\xvec_{j}|\thvec_i, \doublehat\nuisvec_{ij}) - \LL(\xvec_{j}|\hat\thvec_{j}, \hat\nuisvec_{j}).
\end{split}
\end{align}
\end{linenomath*}
Both likelihoods are evaluated on the FC pseudoexperiment spectrum, $\xvec_j$, at parameter values which minimize the likelihood function, $\LL$, but they differ in which parameters are allowed to vary in the minimization. The first likelihood is evaluated after a constrained fit where the parameters of interest are fixed to the values used to generate the pseudoexperiment, $\thvec = \thvec_i$, and only the nuisance parameters are varied, denoted by $\nuisvec = \doublehat\nuisvec_{ij}$, analogous to how $\doublehat\nuisvec_{i}$ is determined in the fit to the real data. 
The second likelihood is evaluated after an unconstrained fit in which both $\thvec$ and $\nuisvec$ are varied in order to find the global minimum of $\LL(\xvec_j)$, denoted, $(\hat\thvec_{j}, \hat\nuisvec_{j})$.

The neutrino oscillation parameter space can be degenerate, in particular for \dcp and nuisance parameters like $\theta_{13}$, or for values of \thetamix mirrored around the value which produces maximal \numu disappearance. In order to avoid biases towards a particular region of parameter space, we run multiple fits with different seed values for each FC pseudoexperiment and then take the result with the lowest \LL. 

Between 1000 and 5000 FC pseudoexperiments are generated at each $\thvec_{i}$, where more FC pseudoexperiments are required for the most extreme $p$-values. Furthermore, given the very large number of FC pseudoexperiments that are required in the 3-sigma (and above) regions in order to accurately measure the corresponding small $p$-values, we choose to only perform the profile construction in regions where $\sqrt{\lambda_\mathrm{Wilks}}<20$ for 1-dimensional constraints and $\sqrt{\lambda_\mathrm{Wilks}}<12$ for 2-dimensional constraints.

\begin{figure}
    \includegraphics[width=.48\linewidth]{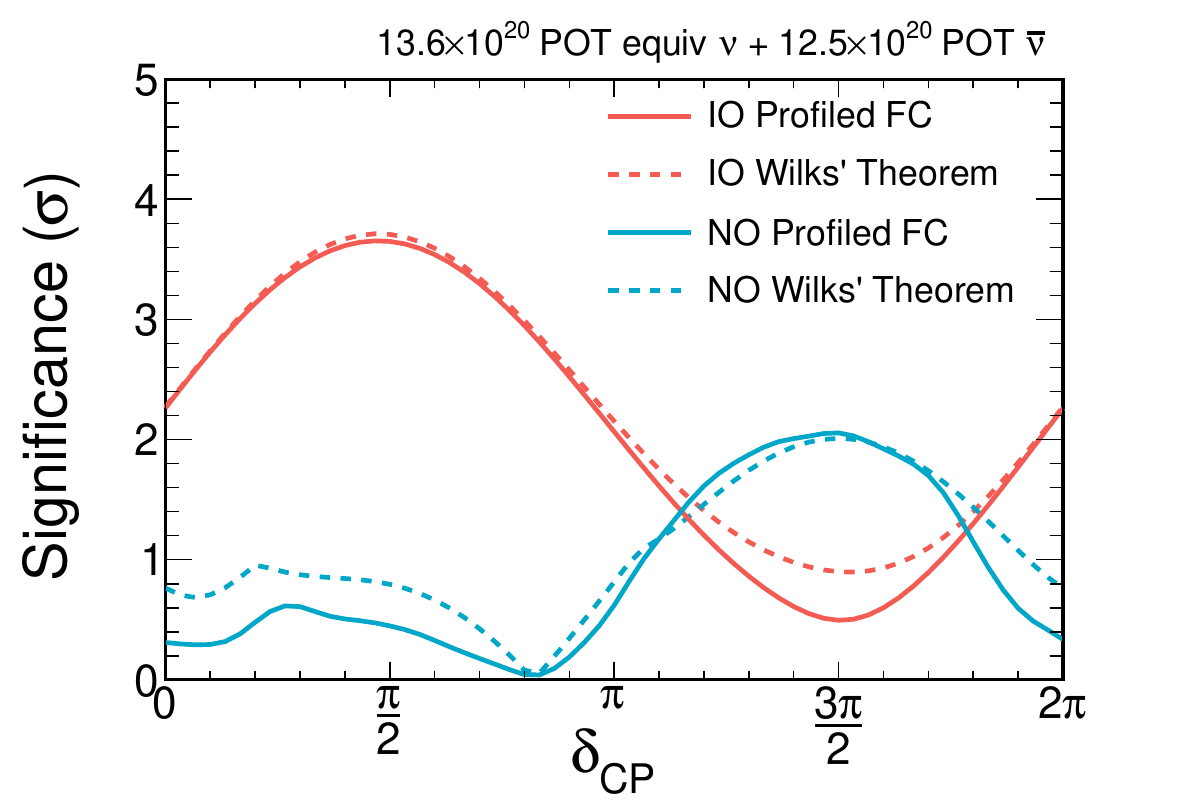}
    \includegraphics[width=.49\linewidth]{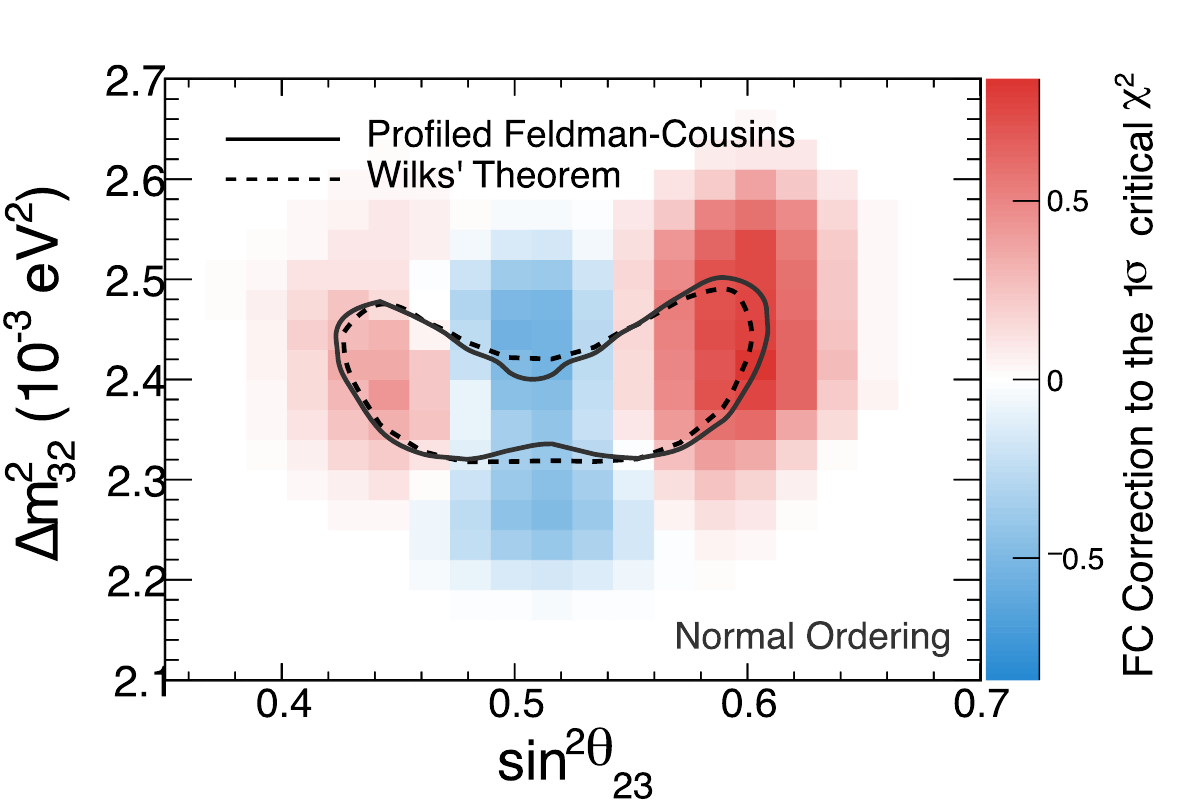}
    \caption{Comparisons between Wilks' theorem (dashed) and the profile construction (solid) for two results from~\cite{ref:nova3f2021}.
    Left: Significance of the data for different values of \dcp and mass ordering. Right: Contour plot showing the 1-$\sigma$ domain of isosignificance in the normal ordering for \dmsq{32} vs.~\sinsq{32}. The right additionally includes a color scale that shows the size of the change in the 1-$\sigma$ critical value at that point in parameter space. The contour `pinches' around \sinsq{23} = 0.51 since that is the point of maximal disappearance, an effective `boundary' in the impact of this parameter.}
    \label{fig:sig_plots}
\end{figure}

The $\LR_{ij}$ distributions are then used to build empirical test statistic distributions for each $\thvec_{i}$. For 1-dimensional significance plots, a $p$-value is first determined at each grid point by counting the fraction of FC pseudoexperiments with a $\LR_{ij}$ larger than that of the data at that same $\thvec_{i}$. The $p$-value is then converted to a significance via $\sigma=\sqrt{2}\erfc^{-1}(p)$. The resulting collection of significances is then interpolated and smoothed taking care to preserve real discontinuous features (discussed more in \sect{limitations}). Figure~\ref{fig:sig_plots} illustrates how significances for one or two parameters of interest can be represented. For most regions of the parameter space, we expect the underlying likelihood surface to be well--behaved but the existence of boundaries and local, nearly degenerate minima can skew the test statistic distributions,  resulting in jump of significances between neighboring grid points, as illustrated in Section~\ref{sec:limitations}.

The procedure to establish 2-dimensional contours of isosignificance is slightly different. We first start by evaluating the standard likelihood of the data at each point $\thvec_{i}$ of the grid used to sample the parameter space. We then evaluate the critical likelihood corresponding to each of the significance levels of interest, namely 1$\sigma$, 2$\sigma$, and 3$\sigma$, from the set of Feldman--Cousins pseudoexperiments, again, at each grid point. Each map of critical profile construction values is then subtracted from the map of standard likelihood obtained from the data. The intersection of the resulting surfaces with the plane 0 (or, for the inverted ordering, with the plane $\LR_{IH}$, which is the difference between the likelihoods of the best fit point in the Inverted Ordering and the overall best fit point) represents the contours of isosignificance. A kernel smoothing procedure is finally applied to the 2-dimensional contours, taking care to consider points near $\dcp=0$ and $\dcp=2\pi$ as neighbors (due to its cyclical nature) in the \sinsq{23} vs.~\dcp contours.

\subsection{Hypothesis tests}\label{sec:hypothesis}

In addition to 1-dimensional and 2-dimensional constraints on oscillation parameters, we can perform hypothesis tests for the mass ordering, the \thetamix octant, or a combination of both. A key benefit of the profile construction is that the procedure can naturally address these binary tests (or discrete choices in general) since when applied to a single point the procedure becomes a classic likelihood ratio test with Monte Carlo used to determine the $p$-value. Similar procedures have been shown in the literature for some time, generally focused on questions of sensitivty of future experiments, for example~\cite{ref:franco}. In our procedure, FC pseudoexperiments are generated with the parameter being tested held fixed and all other parameters set to their profiled values given that constraint. For example, if the overall best fit is in the normal ordering, the test would be for rejecting the inverted ordering, so the FC pseudoexperiments would be generated in the inverted ordering with all other parameters set to the best fit to the data in that ordering.
Since this procedure is only done at one point of the parameter space for each hypothesis test, we can afford to generate more FC pseudoexperiments (tens of thousands) and reach more accurate measurements of the $p$-values and significances than for 1D and 2D confidence intervals. The result of the procedure is, again, an empirical collection of $\LR=\LL_\mathrm{constrained}$-$\LL_\mathrm{unconstrained}$ which can be used to determine the fraction of FC pseudoexperiments that yield a \LR less compatible with the null hypothesis than the data, equating to a $p$-value. This likelihood--ratio test statistic slightly differs from the one defined in \eqn{lrtestprof}: all parameters are still free to vary in the unconstrained fit, but in the constrained fit, the parameters of interest are allowed to take values within the limits defined by the hypothesis being tested. This procedure is the only correct one for the estimation of our level of preference (or rejection) for a given hypothesis; it cannot be done by reading the minima of the 1-dimensional or 2-dimensional confidence intervals, as explained in more detail in \sect{limitations}.
The profile construction can also be extended in a straightforward way to also calculate a CLs significance, see Appendix~\ref{sec:cls} for details.

\subsection{Validation}\label{sec:validation}

When considering any frequentist statistical procedure, a key step is to evaluate the coverage properties of that procedure for the problem at hand. The goal of the profile construction is to produce confidence intervals with coverage as close as possible to the stated level $\alpha$. The examples in \sect{toy} show that none of the procedures considered produce perfect coverage when certain truth quantities are unknown, but in those examples, the profile construction comes the closest. 

Here we give an in-situ demonstration of achieving these coverage properties with NOvA simulation by generating validation pseudoexperiments at known true values, and evaluating how often those true values would be contained within confidence intervals drawn with the profile construction as well as Wilks' theorem for comparison. In the ideal case, we would expect the 50\% confidence intervals to cover the true point in 50\% of the validation pseudoexperiments. Two true test points were chosen: the overall best fit point from~\cite{ref:nova3f2021}, which is far from boundaries, leading to little expected impact from the profile construction on the significance, and the preferred point if the CP--violating phase was $\dcp=0$ where larger deviations are expected due to parameter degeneracies.\footnote{While this test could be done at any points, these points from the fit to NOvA data were chosen to give concrete, relevant examples.}

We perform the test with one-dimensional confidence intervals in \dmsq{32}, though any parameter (or set of parameters) would work. At each true point, 1000 validation pseudoexperiments are generated. For each pseudoexperiment, $i$, we must determine whether the true parameter value used to generate the pseudoexperiments, $\thvec_0$, would be included within the confidence interval drawn at significance $\alpha$. For both methods, the first step is to perform two fits to determine both the overall best fit point, ($\hat\thvec_i$, $\hat\nuisvec_i$), as well as the preferred set of nuisance parameters when $\thvec$ is constrained to the true value the validation pseudoexperiments were generated at, $(\thvec_0, \doublehat\nuisvec_{0i})$\footnote{In this study the nuisance parameters just include other oscillation parameters; we did not include systematic uncertainties.}. The log-likelihood ratio between these two points is then calculated:
\begin{linenomath*}
\begin{equation}
    \lambda_i = \LL(\thvec_0, \doublehat\nuisvec_{0i}) - \LL(\hat\thvec_i, \hat\nuisvec_i).
\end{equation}
\end{linenomath*}
For Wilks' theorem, determining if the true point would be included in confidence interval $\alpha$ is a simple check if $\lambda_i$ is less than the pre-tabulated critical values, \critw, which are the same for every validation pseudoexperiment. For the profile construction, the critical values, \criti, must be found individually for each validation pseudoexperiment, $i$, using
1000 FC pseudoexperiments generated at $(\thvec_0, \doublehat\nuisvec_{0i})$, true value being tested along with the experiment-by-experiment preferred nuisance parameters per the profile construction. The true value would be included within the profile construction confidence interval if $\lambda_i < \criti$. For both methods, the effective coverage at level $\alpha$ is defined as the fraction of experiments where $\thvec_0$ would be included, i.e. $\lambda_i$ is less than the respective critical value.

\begin{figure}
    \centering
    \includegraphics[width=.45\linewidth]{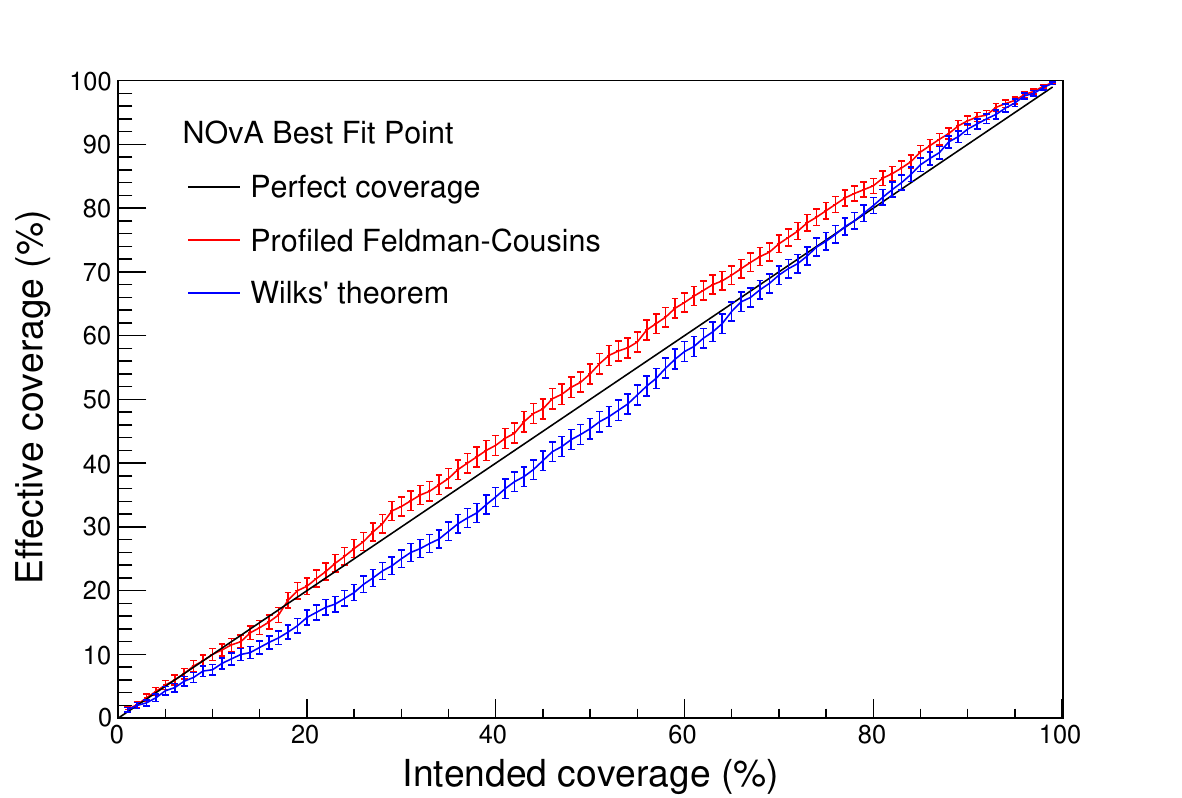}
    \includegraphics[width=.45\linewidth]{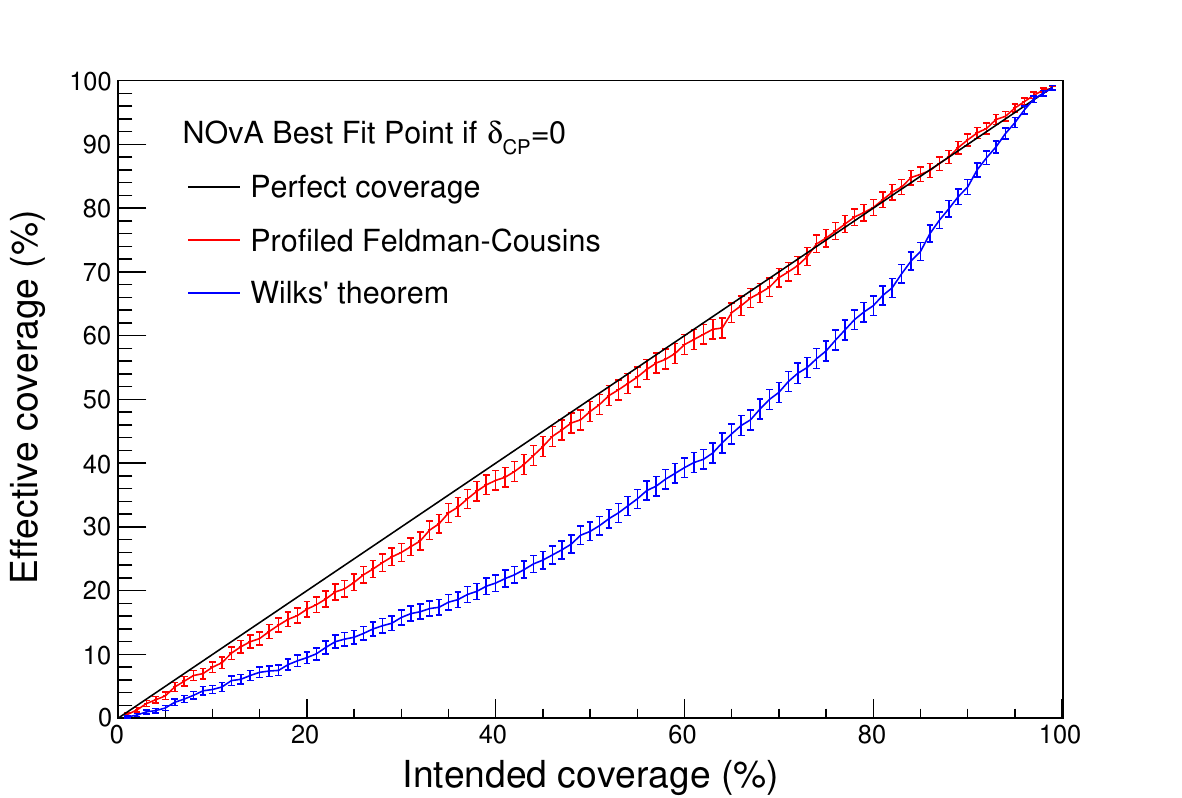}
    \caption{
    The left figure shows the coverages obtained with Wilks' theorem (blue) and the profile construction (red) at our overall best fit point, while the right figure shows those coverages at our best fit if $\dcp=0$.
    On the left, Wilks' theorem shows a good approximate coverage, while on the right, it produces a significant under-coverage, which would have the effect to artificially disfavor $\dcp=0$. The coverage obtained with the Profiled Feldman--Cousins approach is consistently more accurate. The error bars represent the statistical uncertainty on the binomial confidence interval obtained from 1000 fake experiments.}
    \label{fig:cov_comparison}
\end{figure}

Note that without nuisance parameters, this test would be tautological: the validation pseudoexperiments and the FC pseudoexperiments being used to determine if the test point would be inside the profile construction confidence interval would all be drawn based solely on $\thvec_0$, and so the coverage must be correct. In the presence of nuisance parameters, however, the validation pseudoexperiments are drawn based on $(\thvec_{0}, \nuisvec_0)$ while the FC pseudoexperiments are drawn from $(\thvec_0, \doublehat\nuisvec_{0i})$. \fig{cov_comparison} shows how the coverages obtained under Wilks' theorem and the Profiled Feldman--Cousins approach vary for different intended coverages at the two points of parameter space considered above. Wilks' theorem generates widely different results depending on the region of the parameter space and can significantly deviate from the ideal coverage. The Profiled Feldman--Cousins method provides us with a more consistently accurate estimation of the desired coverage. \fig{cov_comparison} hints that the magnitude of the corrections might decrease in the most extreme significance levels. This is not a general property and is further investigated in \sect{limitations}.
We also performed a cross-check of the significance of our mass ordering determination using an alternative (and more conservative) method of handling nuisance parameters developed by Berger and Boos~\cite{ref:bb}. That procedure did not uncover a larger $p$-value than the one reported from the profile construction, and so is consistent with that result. The details of this cross-check can be found in Appendix~\ref{sec:bergerboos}.

\subsection{Limitations and Features}\label{sec:limitations}

The nominal output of the Feldman--Cousins method is a single confidence interval or region with reasonable coverage. However, it is straightforward and convenient to apply a Feldman--Cousins correction to a whole likelihood surface: each point has a likelihood, from that likelihood a $p$-value can be determined based on the distribution of FC pseudoexperiments at that point, and then from that $p$-value work backwards to an equivalent likelihood. This \term{Wilks' Surface} is quite practical to work with since contours at any significance can be drawn using the Wilks' critical values. However, while the Wilks' Surface superficially resembles an actual likelihood, it does not have the properties of a likelihood. Notably, it cannot be `profiled' to reduce its dimensionality: a two-dimensional likelihood surface and its associated FC pseudoexperiments cannot be used to find one-dimensional confidence intervals.

The determination of the mass ordering in the most recent NOvA results provides a clear demonstration of this phenomenon~\cite{ref:nova3f2021}. The lowest significance for the Inverted Ordering has several different values in different projections of the significance: $0.6\sigma$ vs.~\sinsq{23} and $0.5\sigma$ vs.~\dmsq{32} or \dcp. Mechanically, these differ since each projection is determined with different sets of experiments generated at different assumed true values. They are not expected to correspond in principle because assigning the likelihood of the Inverted Ordering as a whole to the lowest value of the likelihood when projected against another variable is an example of profiling, which is not a valid operation on these Wilks' Surfaces. The correct procedure is to generate FC pseudoexperiments specific to each question being asked, in this case a hypothesis test to determine the ordering. A benefit of the FC approach is that it can naturally accommodate binary questions like the neutrino mass ordering where the number of degrees of freedom for the Wilks' theorem approach is not well-defined, typically producing stronger constraints than applying Wilks' theorem with 1 degree of freedom. In this case, the significance calculated directly for rejecting in the Inverted Ordering is $1.0\sigma$.

\begin{figure}
    \newcommand{\discontdcpsubfig}{a}
    \newcommand{\discontprofsubfig}{b}
    \newcommand{\discontpvaluesubfig}{c}
    \centering
    \begin{minipage}{.32\textwidth}
     \includegraphics[width=\textwidth]{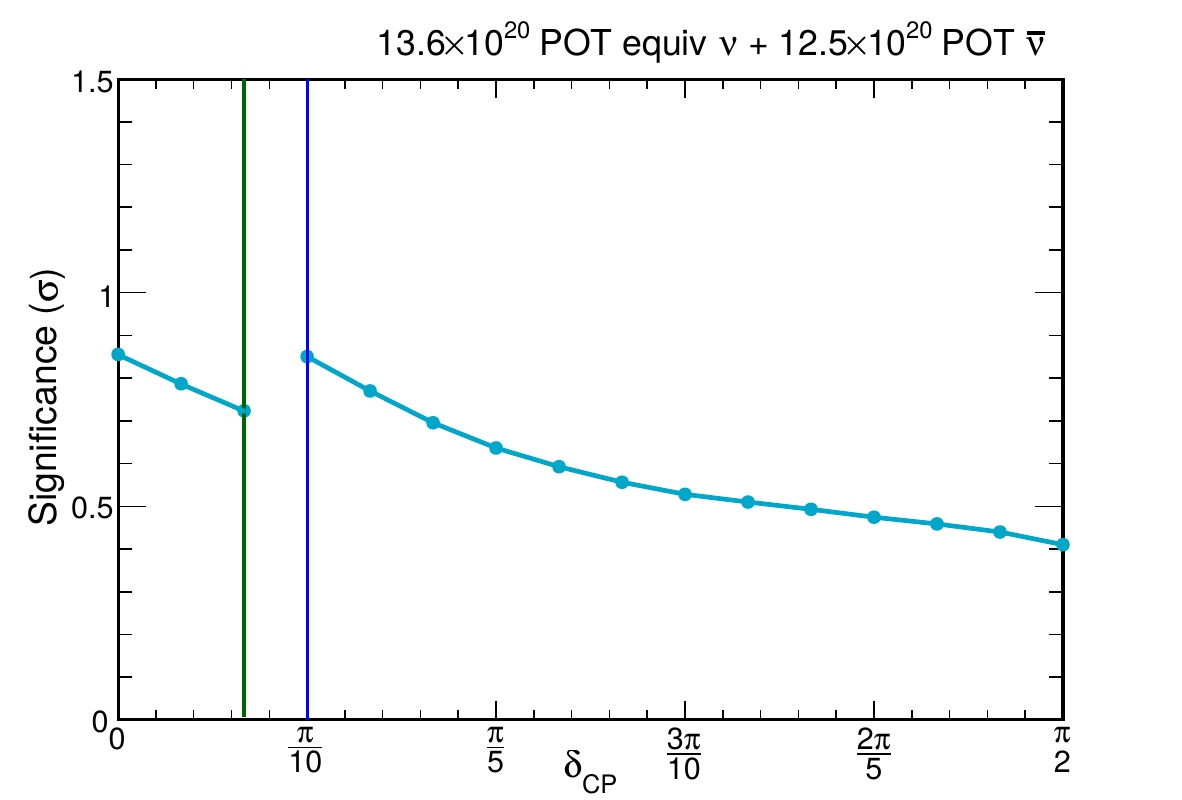}\\
        \footnotesize (a) Significance vs.~\dcp
    \end{minipage}
    \hfill
    \begin{minipage}{.32\textwidth}
        \includegraphics[width=\textwidth]{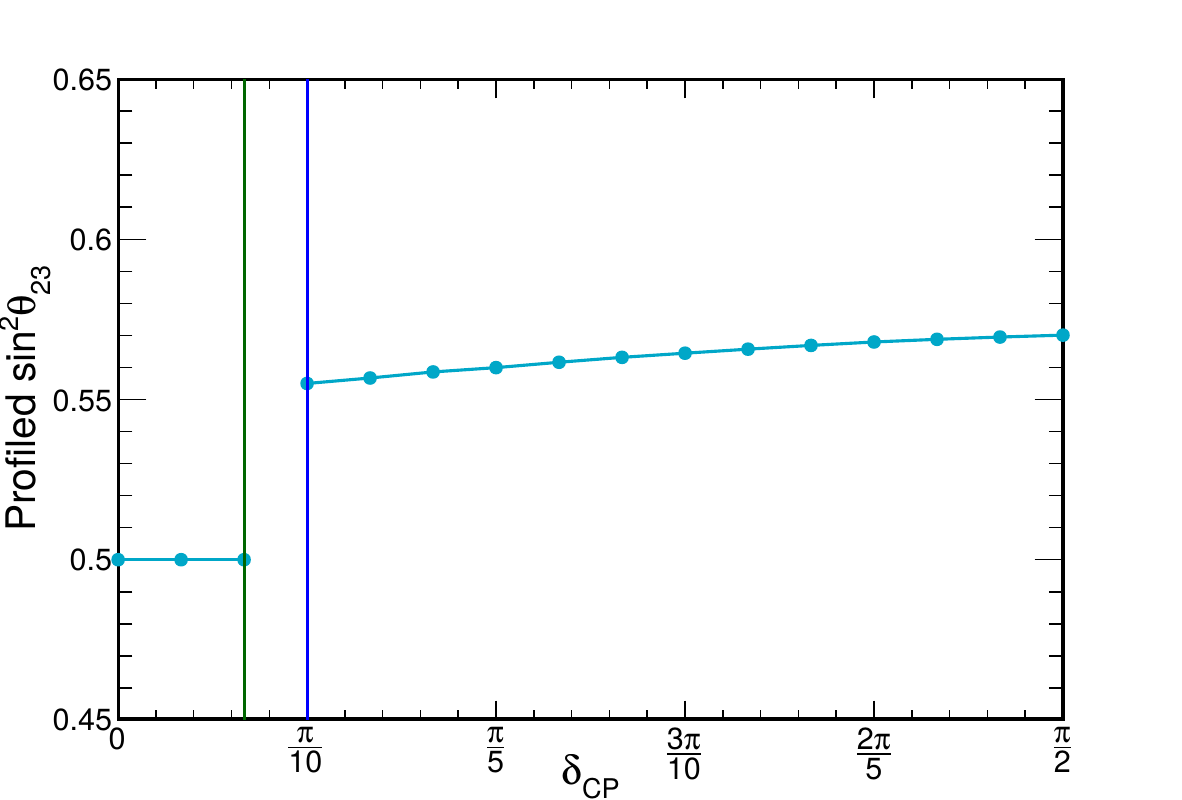}\\
        \footnotesize (b) Profiled \sinsq{23} vs.~\dcp
    \end{minipage}
    \hfill
    \begin{minipage}{.32\textwidth}
        \includegraphics[width=\textwidth]{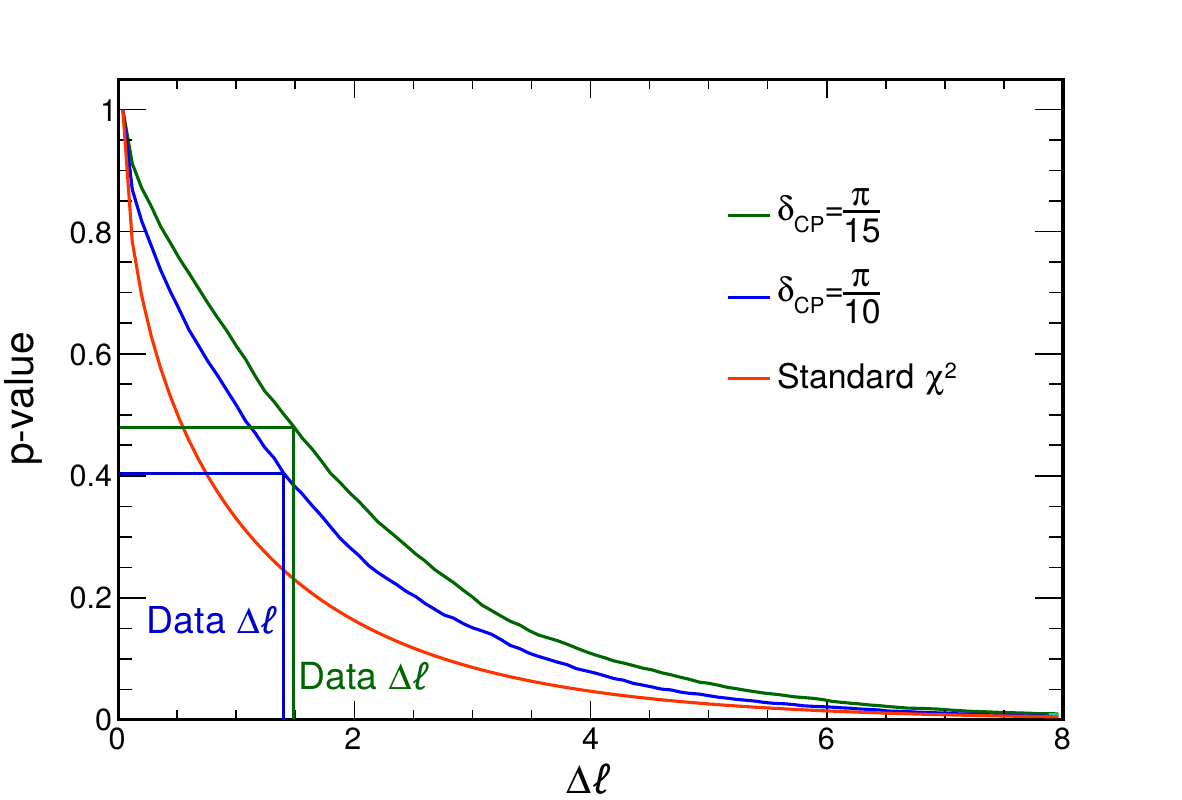}\\
        \footnotesize (c) $p$-value vs.~critical $\LR$
    \end{minipage}

    \caption{(\discontdcpsubfig) The quoted significance vs.~\dcp is discontinuous around $\dcp=\frac{\pi}{10}$. This is due to the discontinuity in the profiled value of $\sinsq{23}$ as a function of $\dcp$. (\discontprofsubfig) \sinsq{23} transitions from maximal mixing to upper octant at this point. The FC pseudoexperiments are therefore generated at different points in parameter space. (\discontpvaluesubfig) The very similar values of $\lambda$ in the data are assigned different p-values due to being compared to different empirical distributions. The $p$-value is obtained by integrating the empirical test-statistic distribution, $P(\lambda)$, from a lower bound, shown here on the x-axis, to $+\infty$.
    }
    \label{fig:disconts}
    \newcommand{\discontpvalue}{\ref{fig:disconts}\discontpvaluesubfig}
    \newcommand{\discontprof}{\ref{fig:disconts}\discontprofsubfig}
    \newcommand{\discontdcp}{\ref{fig:disconts}\discontdcpsubfig}
\end{figure}

With this method, it is also possible for discontinuities in the corrected significance plot to emerge even if the underlying likelihood surface is smooth. An example of one of such a discontinuity can be found in Figure~\ref{fig:disconts}a around $0.1\pi$ in the plot of significance vs.~\dcp in the normal ordering, upper octant. This occurs because of a discontinuity in the profile construction corrections, caused by a discontinuous change in the value of the nuisance parameters\footnote{Discontinuous changes in the nuisance parameters when testing a continuous set of values of a parameter of interest are quite common and typically not a problem. Without FC corrections, these changes can cause a discontinuous change in the \emph{derivative} of the likelihood, but do not make the value of the likelihood discontinuous.}.
In this particular case, the global minimum moves from maximal mixing to the upper octant at this particular value of \dcp, as shown in Figure~\ref{fig:disconts}b, leading to a change in the underlying \LR distributions on either side of the discontinuity which then translates to different $p$-values for a given critical value, shown in Figure~\ref{fig:disconts}c. 

\begin{figure}
    \includegraphics[width=.32\linewidth]{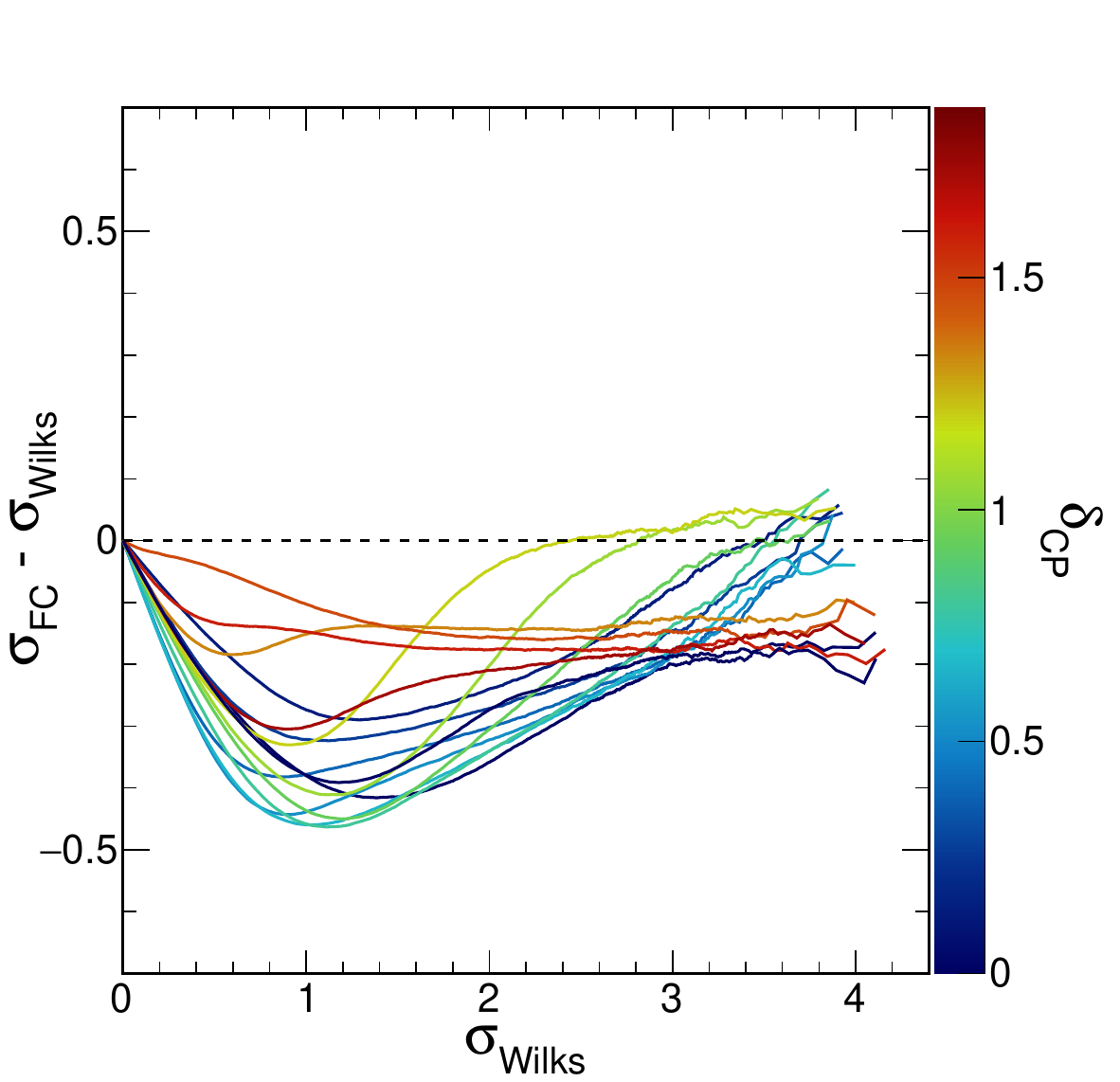}
    \includegraphics[width=.32\linewidth]{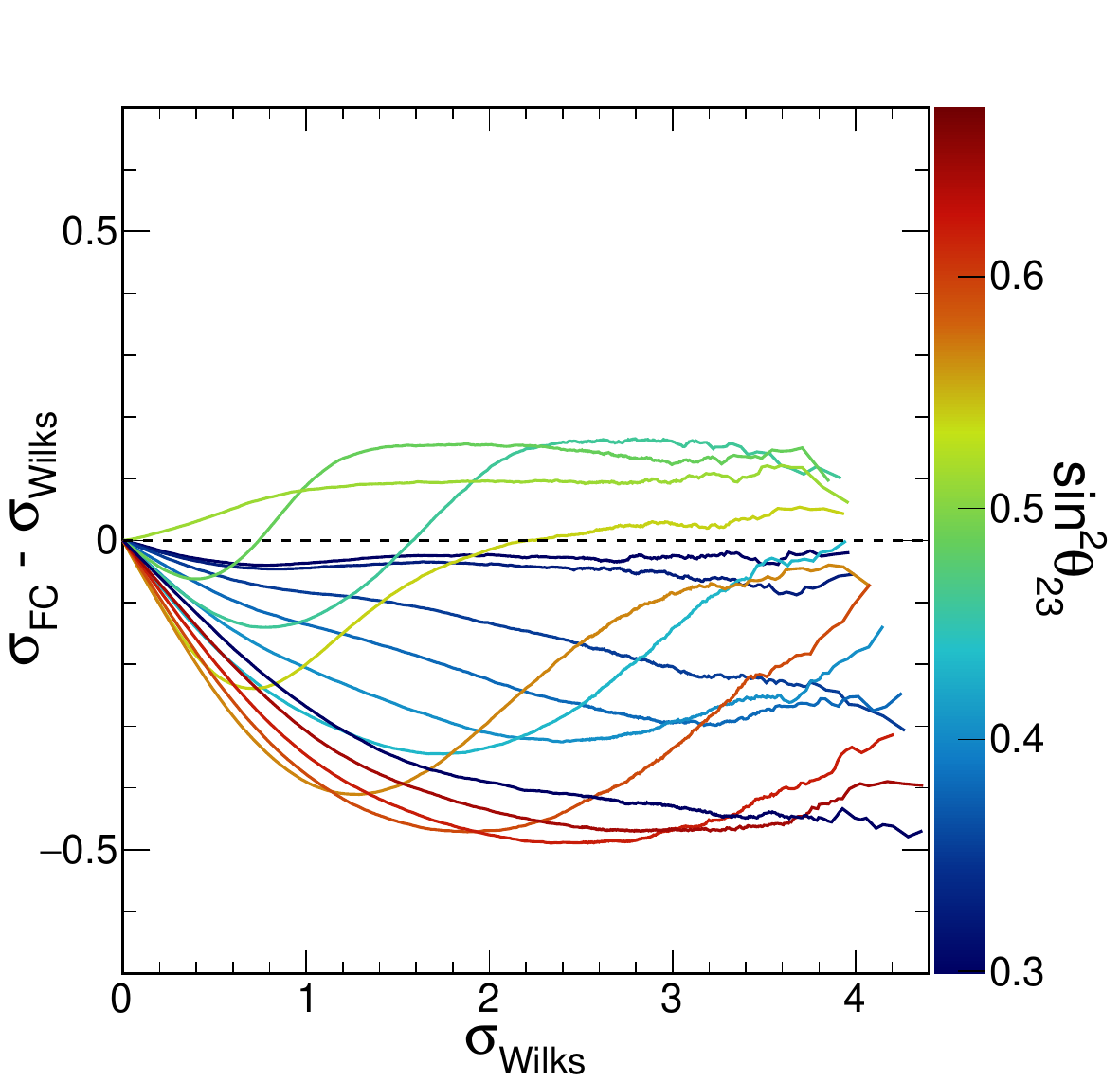}
    \includegraphics[width=.32\linewidth]{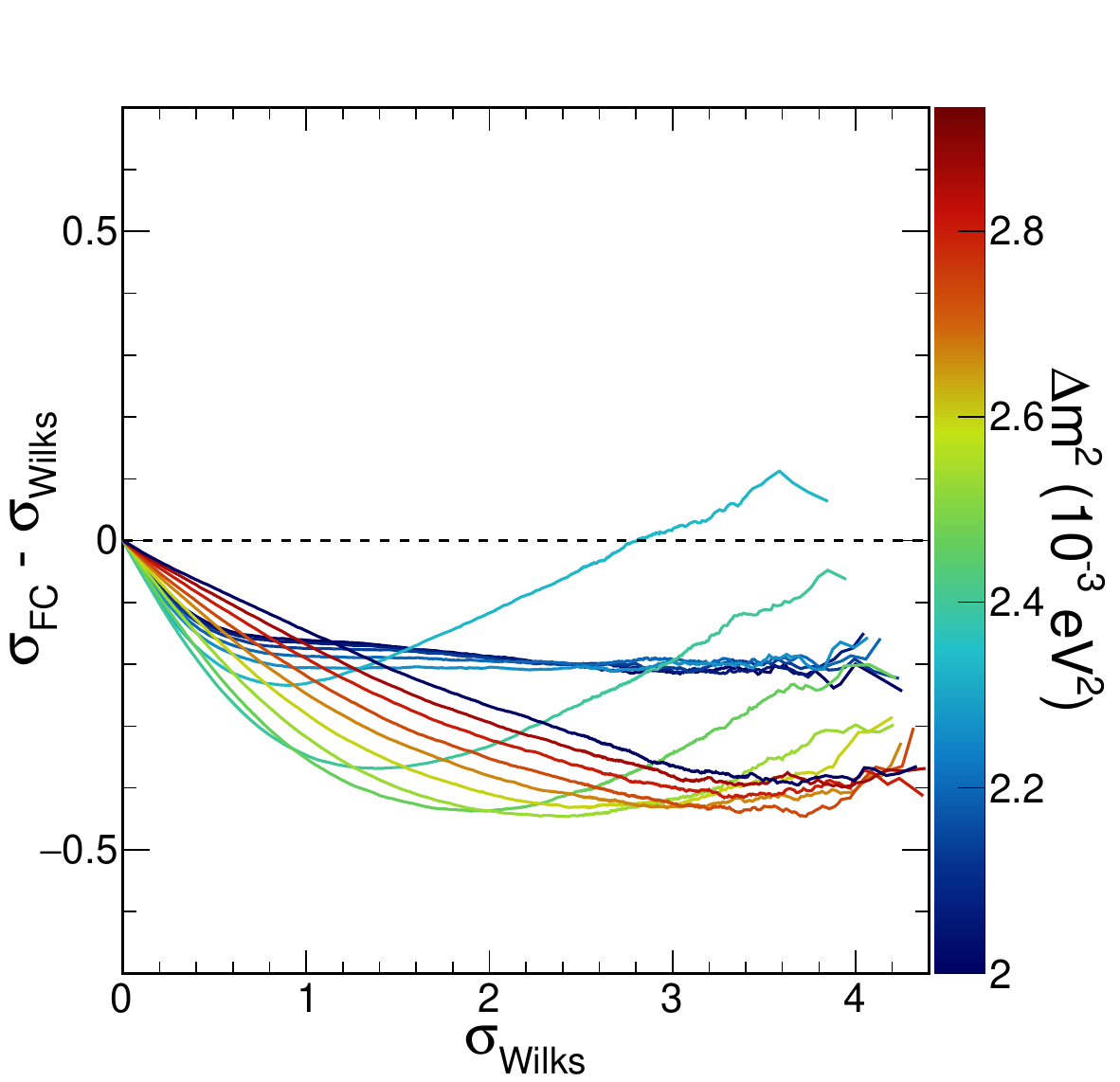}
    \caption{The change in significance vs.~the significance level at which the correction occurs for different values of, from left to right, \dcp, \sinsq{23}, and \dmsq{32}. The colors represent different true values of the parameter in question being tested.}
    \label{fig:highsignificance}
\end{figure}

A drawback of this method is its computation cost. We explored how the size of profile construction corrections depend on the significance for which the correction is being computed. It would be convenient if the size of corrections became smaller as significance increases since corrections require more FC pseudoexperiments and get progressively more expensive to calculate at higher significance. We explored this question using the three 
plots which tested significance for different true values of \dcp, \sinsq{23}, and \dmsq{32}, and the results are shown in \fig{highsignificance}. While the sizes of corrections clearly change as a function of significance, and for some true values the corrections converge towards zero, this is not true in general: the sizes of corrections at $4\sigma$ can be as large as the corrections at $2\sigma$. In these examples, the \emph{relative} size of the correction does decrease as the absolute significance gets larger, but we leave it to the reader to decide if the difference between $3.5\sigma$ and $4\sigma$ is more or less important than the difference between $1.5\sigma$ and $2\sigma$.

Another limitation is that it is not possible to combine the corrected likelihoods from two separate experiments to produce a combined likelihood surface from a joint analysis. While it is possible to combine experiments using FC corrections, doing so requires more detailed information than is captured in just the likelihood and corrections~\cite{ref:alfons}.

\section{Conclusions}

The Feldman--Cousins method provides a method for handling the common challenges that experiments encounter when Wilks' theorem cannot be relied upon, but the lack of a prescription for handling nuisance parameters complicates its adoption in practice. 
The NOvA experiment has adopted the profile construction for its oscillation measurements~\cite{ref:nova3f2017, ref:nova3f2018, ref:nova3f2019, ref:nova3f2021}, which offers a straightforward prescription for handling nuisance parameters. 
Toy studies inspired by these oscillation measurements show the method achieves more accurate coverage when the true parameters of the underlying model are unknown compared to other plausible methods, and toy studies with constrained systematic uncertainties show similar performance to other methods. In-situ tests in the NOvA analysis further validate the accuracy of the reported confidence intervals and significances.
The most significant challenge to making use of profile construction (and Feldman--Cousins in general) is the large computational cost associated with generating and fitting the required FC pseudoexperiments. Our approach takes advantage of available High Performance Computing resources, but other approaches to improve the efficiency of this method are also being explored~\cite{ref:gausprocess,ref:berns}.

\section{Acknowledgments}
This document was prepared by the NOvA collaboration using the resources of the Fermi National Accelerator Laboratory (Fermilab), a U.S. Department of Energy, Office of Science, HEP User Facility. Fermilab is managed by Fermi Forward Discovery Group, LLC, acting under Contract No. 89243024CSC000002. This work was supported by the U.S. Department of Energy; the U.S. National Science Foundation; the Department of Science and Technology, India; the European Research Council; the MSMT CR, GA UK, Czech Republic; the RAS, MSHE, and RFBR, Russia; CNPq and FAPEG, Brazil; UKRI, STFC and the Royal Society, United Kingdom; and the state and University of Minnesota. This research used resources of the National Energy Research Scientific Computing Center (NERSC), a Department of Energy Office of Science User Facility using NERSC award to project m3990 in 2020-22. We are grateful for the contributions of the staffs of the University of Minnesota at the Ash River Laboratory, and of Fermilab. 
\appendix

\section{CLs Mass Ordering Significance}\label{sec:cls}

The \CLs method~\cite{ref:cls1,ref:cls2,ref:cls3} was introduced as an alternative to traditional $p$-value calculations to address situations where an experiment might potentially make a claim of \term{discovery} well beyond its sensitivity. In a nutshell, the method takes a ratio between the $p$-value for the null hypothesis, $\mathcal{H}^0$, and the potential discovery hypothesis, $\mathcal{H}^1$. In a true discovery, $p(\mathcal{H}^0) \ll p(\mathcal{H}^1)$, and the \CLs value will be small, while in a spurious claim, the data will be a poor fit to both hypotheses, so even though $p(\mathcal{H}^0)$ might be small, \CLs will be of order 1.

\begin{figure}[bth]
    \centering
    \includegraphics[width=.75\linewidth]{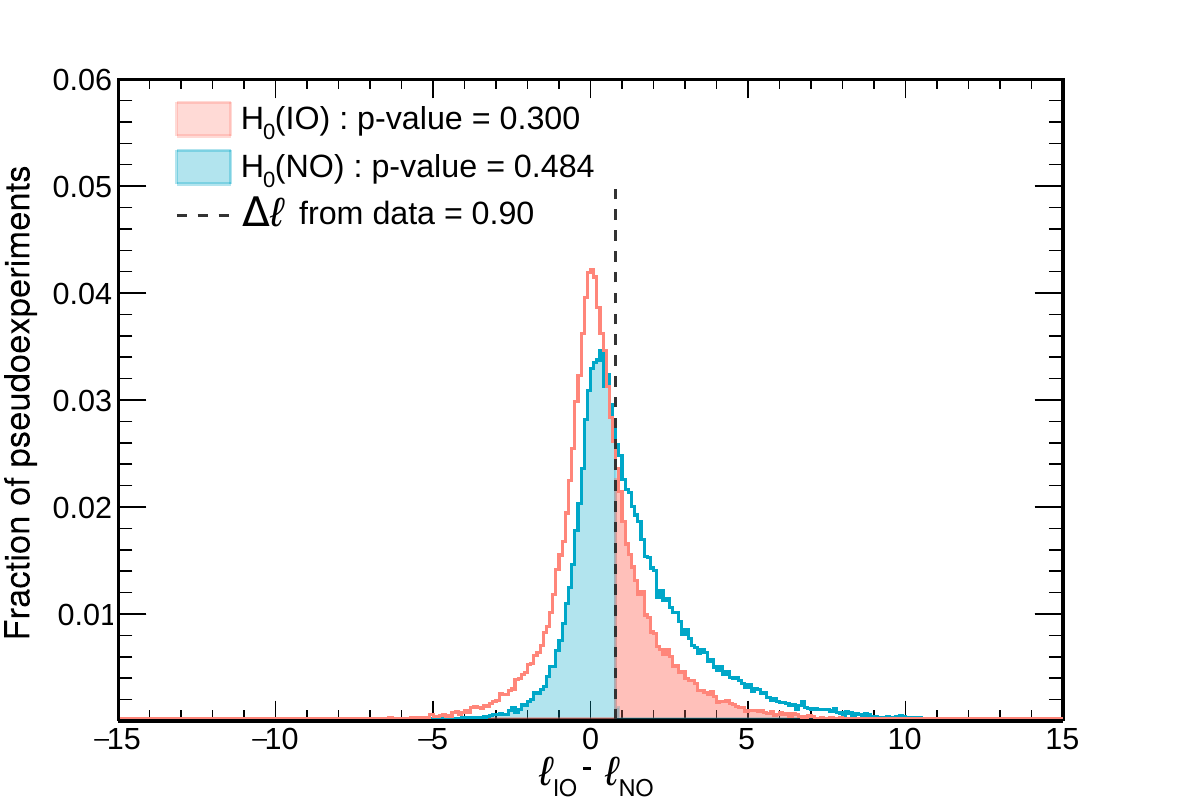}
    \caption{Distribution of the likelihood ratio $\LR=\ell_\mathrm{IO}-\ell_\mathrm{NO}$ for FC pseudoexperiments generated at the best fit points in the IO (red) and the NO (blue). The fraction of FC pseudoexperiments with a likelihood ratio more compatible with the null hypothesis than the data is smaller in the case of the NO, which suggests a preference for the latter. The resulting \CLs factor is 0.620.}
    \label{fig:CLs}
\end{figure}

In the particular case of binary questions, the profile construction can be naturally extended so the same FC pseudoexperiments can be re-used for the \CLs method . A mass ordering test is presented here, but the method is generic. Two modifications are needed. First, rather than evaluating $\LL_\mathrm{constrained}$ and $\LL_\mathrm{unconstrained}$, $\LL_\mathrm{NO}$ and $\LL_\mathrm{IO}$ are evaluated, but they can be readily re-interpreted: $\LL_\mathrm{constrained}$ corresponds to the $\LL$ for the hypothesis being tested and $\LL_\mathrm{unconstrained}$ corresponds to whichever $\LL$ is lower\footnote{Since FC pseudoexperiments generated in the Normal Ordering may have a better fit in the Inverted Ordering, and vice versa, these two $\LL$'s may be the same or not.}. Second, FC pseudoexperiments need to be generated for both possible hypotheses, but given the relatively low computational cost of this test, this is a minor overall additional cost. Where the profile construction only reports the fraction of FC pseudoexperiments in the hypothesis being tested with $\LR$ larger than that observed in data, \CLs also requires the \term{inverse}: the fraction of FC pseudoexperiments generated under the hypothesis favored by the data with $\LR$ \emph{lower} than that observed in the data, as shown in \fig{CLs}.  A small overlap of the two distributions would signify a strong discrimination power towards the mass ordering. Our data suggests a slight preference for the Normal Ordering.

\section{Validation of Significance in Mass Ordering Determination}\label{sec:bergerboos}

In the case of binary questions, like the choice of ordering, the situation is better thought of as a hypothesis test than a confidence interval, though they are closely related as described in \sect{methods}. For these cases, there is an alternative approach to handling nuisance parameters developed by Berger and Boos~\cite{ref:bb}. In this procedure, the $p$-value of a set of parameter values being tested, \thvec, is redefined as:
\begin{linenomath*}
\begin{equation}
    p_{\textrm{BB}}(\thvec) = \max_{\nuisvec} p(\thvec, \nuisvec) + \beta,
\end{equation}
\end{linenomath*}
where the $\max$ represents the largest $p$-value over all values of the nuisance parameters, \nuisvec, allowed at the $\beta$ confidence level based on a fit to the data. By contrast, the Profiled Feldman--Cousins approach simply uses the $p$-value at $\doublehat \nuisvec$, the maximum likelihood estimate of the nuisance parameters given $\thvec$:
\begin{linenomath*}
\begin{equation}
    p_{\textrm{FC}}(\thvec) = p(\thvec, \doublehat\nuisvec),
\end{equation}
\end{linenomath*}
effectively assuming that the nuisance parameters  which give the largest likelihood value (and thus the largest $p$-value under Wilks' theorem) will also have the largest $p$-value with the pseudoexperiment--calculated critical values. The Berger--Boos method is more conservative since it allows for the possibility that a seemingly non-optimal set of nuisance parameters will produce a \term{favorable} change in the critical value and thus produce a larger effective $p$-value, but it is commensurately more costly to calculate since pseudoexperiments must be produced for a range of nuisance parameters. 
\begin{figure}
    \centering
    \includegraphics[width=.45\linewidth]{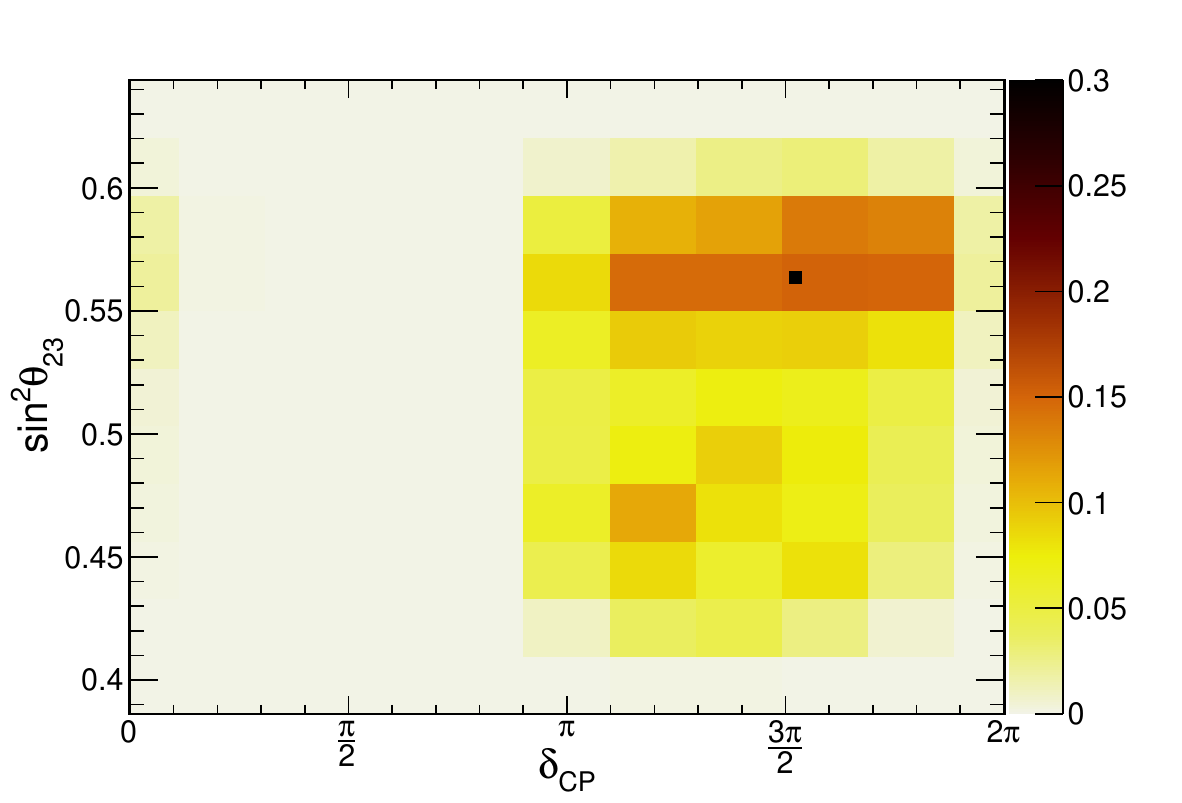}
    \includegraphics[width=.45\linewidth]{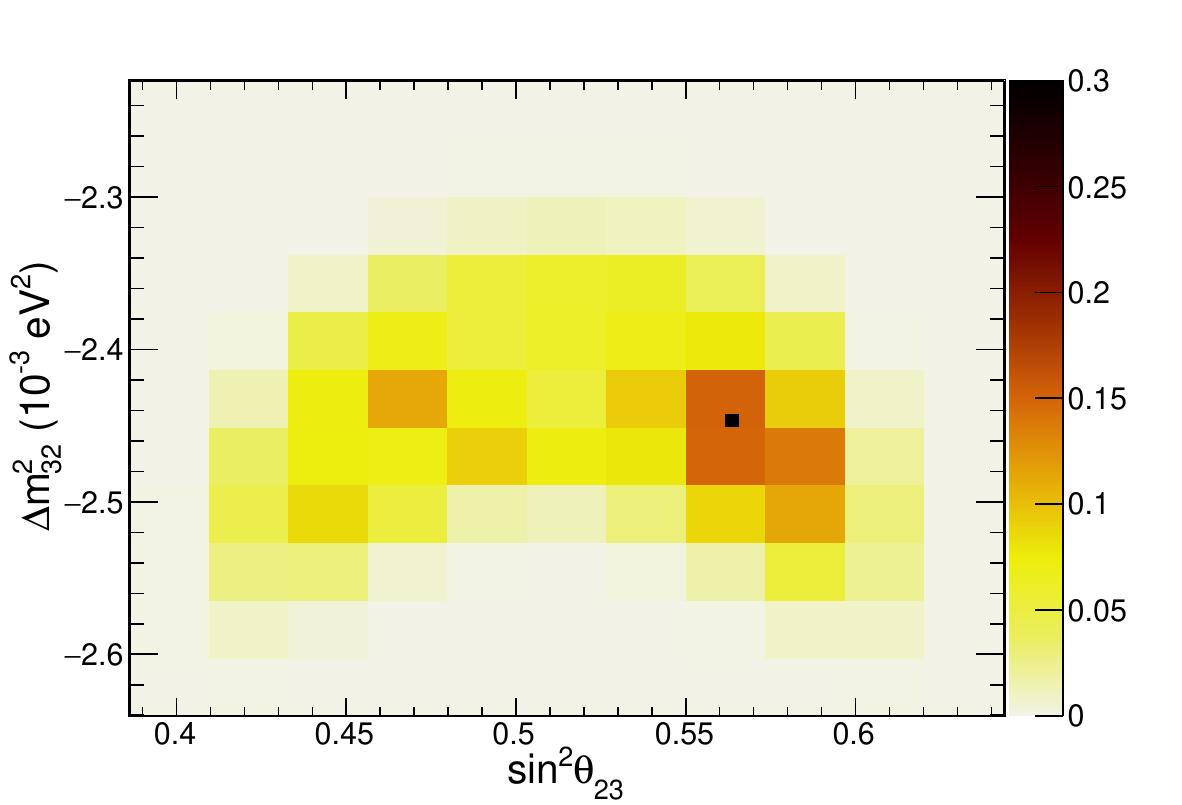}
    \caption{The maximum $p$-values for the tested choices of nuisance parameters in the Berger--Boos test. All points in the full 3-dimensional space were tested, but only the largest $p$-value for each pair of values of the nuisance parameters is shown. All values are below $p=0.30$, the maximum of the color scale and the significance of rejecting the inverted ordering at the best fit point, shown with a small square.}
    \label{fig:bergerboos}
\end{figure}

In practice, it is not possible to test \term{all} values in a multi-dimensional parameter space without an analytic form, so the possible choices of nuisance parameters must be sampled in a fashion which covers the possible space, and for each sampled set of nuisance parameters, a set of FC pseudoexperiments must be generated and used to calculate a new $p$-value. In this case, we are testing the $p$-value for rejecting the IO from the fit to data, $p=0.30$~\cite{ref:nova3f2021}, so are taking a $\beta$ of 0.005 which would not qualitatively alter the interpretation of the original $p$-value. This value of $\beta$ then defines the ranges over which values of the nuisance parameters need to be sampled: a range in \dmsq{32} of $[-2.623, -2.241] \times 10^{-3}\, \textrm{eV}^2$, a range in \sinsq{23} of $[0.397, 0.633]$ and all values of \dcp. Then, 1331 choices of nuisance parameters were tested (11 values in each dimension), sampled uniformly from the possible space, and $p$-values were calculated for those choices. In order to save computational costs, pseudoexperiments were only generated for points where Feldman--Cousins corrections could plausibly raise it above the original $p$-value. The threshold chosen was $\LR < 2.8$, which corresponds to $p_{\textrm{Wilks}} > 0.094$ assuming one degree-of-freedom. A total of 54 points fell below that threshold. 

The largest $p$-value found was $p=0.151$ at $\dmsq{32}=-2.43\times 10^{-3}\, \textrm{eV}^2$, $\sinsq{23}=0.562$, and $\dcp=1.64\pi$, which is below the $p=0.30$ at the best fit point, so the original $p$-value is still the largest. This point had a $\LR = 1.10$, which would give $p_{\textrm{Wilks}} = 0.295$ assuming one degree-of-freedom. This behavior was typical of most points for which FC pseudoexperiments were generated: $p$-values decreased (i.e., significances increased) since a binary question effectively has fewer degrees of freedom than one continuous parameter. Only 2 of the 54 points tested had $p > p_{\textrm{Wilks}}$, namely $p = 0.150$ and $p = 0.134$. The plots in \fig{bergerboos} show the largest $p$-values for rejecting the inverted ordering for different choices of the nuisance parameters.

\end{document}